%% file: main.tex
\documentclass[acmsmall]{acmart}

\AtBeginDocument{%
  \providecommand\BibTeX{{%
    \normalfont B\kern-0.5em{\scshape i\kern-0.25em b}\kern-0.8em\TeX}}}

\usepackage{color}
\usepackage{macro}
\usepackage{subfigure,enumitem}
\usepackage{natbib}

\allowdisplaybreaks

\copyrightyear{2020}
\setcopyright{acmcopyright}
\acmJournal{POMACS}
\acmYear{2020} \acmVolume{4} \acmNumber{2}
\acmArticle{29} \acmMonth{6} \acmPrice{15.00}
\acmDOI{10.1145/3392147}
\acmISBN{978-1-4503-7985-4/20/06}

\received{January 2020}
\received[revised]{February 2020}
\received[accepted]{March 2020}

\begin{document}

\newlength{\mywidth}
\setlength{\mywidth}{30em}


\title{Privacy-Utility Tradeoffs in Routing Cryptocurrency over Payment Channel Networks}



\author{Weizhao Tang}
\affiliation{%
  \institution{Carnegie Mellon University, USA}
}
\email{wtang2@andrew.cmu.edu}

\author{Weina Wang}
\affiliation{%
  \institution{Carnegie Mellon University, USA}
}
\email{weinaw@cs.cmu.edu}

\author{Giulia Fanti}
\affiliation{%
  \institution{Carnegie Mellon University, USA}
}
\email{gfanti@andrew.cmu.edu}

\author{Sewoong Oh}
\affiliation{%
  \institution{University of Washington, USA}
}
\email{sewoong@cs.washington.edu}


\renewcommand{\shortauthors}{Tang, et al.}

\begin{abstract}
    Payment channel networks (PCNs) are viewed as one of the most promising scalability solutions for cryptocurrencies today \cite{lightning}. 
    Roughly, PCNs are networks where each node represents a user and each directed, weighted edge represents funds escrowed on a blockchain; these funds can be transacted only between  the endpoints of the edge.
    Users efficiently transmit funds from node A to B by relaying them over a path connecting A to B, as long as each edge in the path contains enough balance (escrowed funds) to support the transaction.  
    Whenever a transaction succeeds, the edge weights are updated accordingly.
    In deployed PCNs, channel balances (i.e., edge weights) are not revealed to users for privacy reasons; users know only the initial weights at time 0.
    Hence, when routing transactions, users typically first guess a path, then check if it supports the transaction. 
    This guess-and-check process dramatically reduces the success rate of transactions. 
    At the other extreme, knowing full channel balances can give substantial improvements in transaction success rate at the expense of  privacy.  
    In this work, we ask whether a network can reveal \emph{noisy} channel balances to trade off privacy for utility.  
    We show fundamental limits on such a tradeoff, and propose noise mechanisms that achieve the fundamental limit for a general class of graph topologies. 
    Our results suggest that in practice, PCNs should operate either in the low-privacy or low-utility regime; it is not possible to get large gains in utility by giving up a little privacy, or large gains in privacy by sacrificing a little utility. 
\end{abstract}



\maketitle

\input{sections/1_introduction.tex}

\input{sections/2_model.tex}
\input{sections/3_tradeoff.tex}
\input{sections/4_clique.tex}
\blue{
\input{sections/5_utility_metrics.tex}
}
\input{sections/6_numerical.tex}

\input{sections/7_related.tex}

\input{sections/8_discussion.tex}

\begin{acks}
    This work is supported in part by NSF grants CIF-1705007 and CCF-1927712, ARO grant W911NF-17-S-0002, Input Output Hong Kong, the Franz Family Fund, and IC3. The authors would like to thank Kuang Xu for shepherding our paper, as well as the anonymous reviewers for their thoughtful comments. 
    \revise{The authors would also like to thank Zhichun Lu for helpful feedback. }
\end{acks}

\bibliographystyle{plain}
\bibliography{references}

\apdx{
\newpage
\appendix

\section*{Appendix}

\input{sections/appendix.tex}
}

\end{document}

%% file: sections/1_introduction.tex
\section{Introduction} \label{sec:intro}
As the adoption of cryptocurrencies grows to unprecedented levels \cite{adoption}, the scalability limitations of these technologies have become apparent.
For example, Bitcoin today can process up to seven transactions per second with a confirmation latency of hours \cite{eyal2016bitcoin,ohie}. 
For comparison, the Visa network can process tens of thousands of transactions per second with a confirmation latency of seconds \cite{visa}.
This gap raises questions about whether cryptocurrencies are fundamentally able to support as much traffic as traditional, centralized solutions. 

In response to these challenges, several cryptocurrencies have turned to a class of scalability solutions called \emph{payment channel networks} (PCNs) \cite{lightning}. 
The core idea is that instead of committing every transaction to the blockchain, a separate overlay network (the PCN) is maintained, in which each node represents a user, and each edge represents pre-allocated funds that can be efficiently and quickly transacted between the two endpoints of the edge under a mutual agreement. 
Critically,  those transactions on the PCN are committed to the blockchain only in periodic batches, which reduces the frequency with which users must call the (slow, inefficient) blockchain consensus mechanism. 
Users can send money to non-adjacent nodes on the PCN by relaying money through intermediate  nodes on the graph.
PCNs are viewed in the blockchain community as one of the most promising scalability solutions today \cite{maxwell,lightning_news}, and several major cryptocurrencies are staking their long-term scaling plans on PCNs. 
Prominent examples include Bitcoin's Lightning network \cite{lightning} and Ethereum's Raiden network \cite{raiden}.

Despite this excitement, there remain several  
technical challenges. 
Principal among them is a privacy-preserving routing problem: 
each time users wish to route a transaction, they must find a path through the PCN with enough pre-allocated funds to route the transaction. 
However, in today's PCNs, edge balances are not publicly revealed for privacy reasons, making it difficult to find such a route reliably. 
The goal of this paper is to study privacy-utility tradeoffs that arise in such a PCN transaction routing. 
Before defining the problem more precisely,  we begin with a brief primer.

\subsection{Payment Channel  Networks (PCNs)} \label{sec:intro:bg}


    
    

    
    

A \emph{payment channel}  is a transaction between two parties that escrows currency for use only between  those two  parties  for some amount of time. 
For example, Alice and Bob could escrow 4 and 2 tokens, respectively, for the next week.  
The escrow transaction is committed to the blockchain to finalize it. 
Once  the channel is finalized, Alice and Bob can send escrowed funds back and forth by digitally  signing the previous state of the channel and the new updated transaction.  
For example, Alice can send 3 of her 4 tokens to Bob, so  that the new channel state is (Alice=1, Bob=5). 
Once the parties decide to close the channel, they can commit its final state through another blockchain transaction. 
Cryptographic and incentive-based protections prevent users from stealing funds in a channel, e.g., by committing an outdated state. 
Maintaining a payment channel has an opportunity cost because users must lock up their funds while the channel is active, and they are not actually  paid until the channel is closed. 
Hence, it is not practical to expect users to maintain a channel with every individual with whom they may ever need to transact.

\begin{figure}
    \centering
    \includegraphics[width=.7\mywidth]{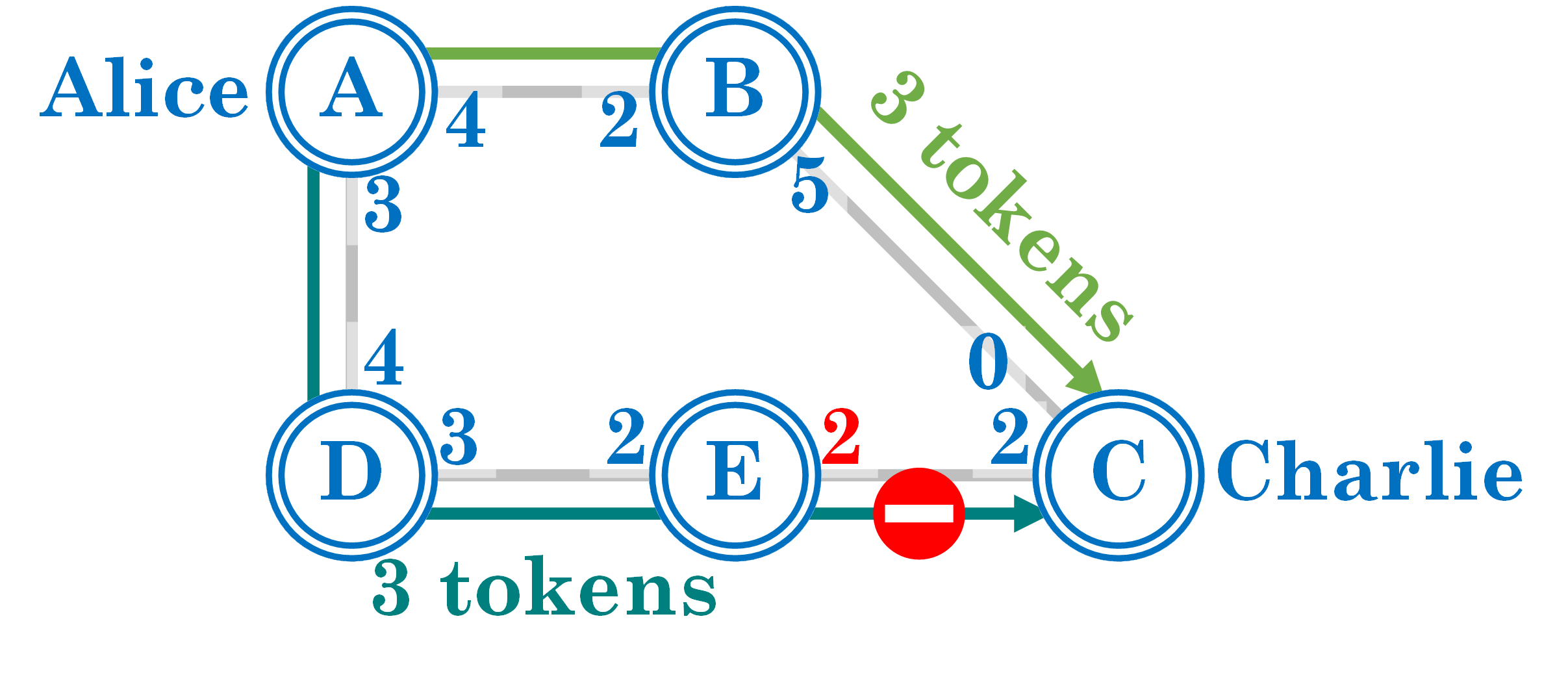}
    \caption{Payment channel network. Alice wants  to send 3 tokens to Charlie.
    Transaction fees are omitted for simplicity.}
    \label{fig:pcn}
\end{figure}

A \emph{payment channel network} (PCN) gets around this problem by setting up a graph of bidirectional payment channels. 
The key idea is that if Alice wants to transact with Charlie, but is only connected to him via Bob, then Bob can act as a relay for Alice's money, passing it along to Charlie. 
Again, cryptographic protections are used to ensure that Bob does not steal Alice's relayed money, but he does receive a small fee as payment for his cooperation.
Notice that if Alice wants to send $r$ tokens to any node in the network,  she must first find a directed path to that node with at least $r$ tokens on every (directed) edge. 
For example, in Figure \ref{fig:pcn}, suppose Alice wants to send $r=3$ tokens to Charlie.
She has two paths available: $A \to B \to C$ and $A  \to D \to E \to C$. 
However, the edge $E \to C$ has only 2 tokens to send in that direction, so it cannot support Alice's transaction.
Hence, this transaction fails. 
Suppose Alice instead sends her transaction over the first path,  $A \to B \to C$.
After her transaction is processed, each of the channels moves $r=3$ tokens (plus a small processing fee) from one side of the channel to the other. 
Note that one could also packet-switch transactions, so Alice sends part of her transaction over multiple paths. 
In today's PCNs, this functionality is not yet implemented \cite{lnd}; however, such routing schemes can give significant  performance enhancements \cite{sivaraman2018routing,dong2018celer,piatkivskyi2018split,amp}.
Packet-switched routing is beyond our scope.

Our work studies the act of finding a path from source to destination  with sufficient balance to support a transaction. 
If the full graph and edge weights were known, finding such a path would  be straightforward (albeit  computationally-intensive). 
However, if users were given access to instantaneous channel balances of the whole network, any passive observer could trivially  see that, for instance, $r$ tokens flowed along a path from Alice to Charlie.  
This is a serious breach of privacy. 

Today's PCNs do not reveal instantaneous balance information in an attempt to prevent observers from inferring other users' transaction patterns. 
Instead, users are given access only to the graph topology, and the \emph{sum} of balances in either direction on each channel, which remains constant for the duration of the channel.  
Because of this design decision, PCN users are forced to guess if a given path has enough balance to support a particular transaction  by attempting to send their transaction  over that path. 
If it fails, the user tries another path until the transaction either completes or times out; upon  a timeout, the user can  instead  process her transaction  on the blockchain, which is comparatively slow and expensive.
This guess-and-check routing approach uses unnecessary resources and severely limits the success rates of today's PCNs \cite{success}. 
Every time a user tries to route over a path that will ultimately fail, it temporarily ties up funds along that path, which cannot be used by other transactions.
Moreover, since users are forced to guess channel balances, they are more likely to not find any valid path prior to transaction timeout. 
Our goal in this work is to understand whether privacy must always come at such a high cost. 
In particular, we consider whether a system could  reveal  \emph{noisy} channel balances in an effort to gracefully trade off privacy for utility. 

\subsection{Contributions}


We make four main contributions:
\begin{itemize}[leftmargin=*]
    \item We theoretically model the routing problem in PCNs and define distribution-free metrics for privacy and utility. 
    \blue{In particular, we relate the \emph{success rate} of a scheme, or the fraction of successfully routed transactions, to a simplified but analytically tractable quantity we call \emph{utility}.}
    %
    
    \item We show a restrictive, so-called \emph{diagonal} upper bound on the privacy-utility tradeoff for these metrics over general graphs and a significant class of shortest-path transaction routing strategies. 
    We show that the diagonal bound is tight by designing noise mechanisms that achieve it. 
    
    \item The diagonal bound is a somewhat negative result, suggesting that a good tradeoff is not possible. 
    However, we show that by relaxing certain assumptions (e.g., the shortest-path routing assumption), we can break the diagonal barrier.
    Indeed, one can design noise mechanisms that asymptotically achieve a perfect privacy-utility tradeoff. 
    However, this comes at the cost of increasingly long paths, i.e., increasingly expensive routing fees.
    
    \item  We demonstrate through simulation that even if one were to consider an average-case utility metric (fraction of successful transactions, or success rate) rather than a worst-case one, the privacy-success rate tradeoff is still not favorable for shortest-path routing. 
    %
    
    %
    Overall, our simulations suggest that trading off privacy for utility does not give significant gains unless the system operates either in a low-privacy regime or low-utility regime; today's PCNs operate in the low-utility regime. 
\end{itemize}

In sum, our results suggest that PCNs may not be able to provide utility and privacy simultaneously.
Moreover, our theoretical analysis is conducted under an  adversarial model that  (a) is passive, and (b) does not exploit temporal correlations in transaction patterns. 
Hence, actual privacy threats are likely even more dire than  our results indicate. 
PCN system designers may therefore need to make an  explicit choice regarding whether the value of PCNs comes mainly from their potential for improving performance or privacy, and choose an operating point accordingly. 


%% file: sections/2_model.tex
\section{Model} \label{sec:model}
We model the PCN as a graph $\mathcal G(\mathcal V, \mathcal E)$, where $\mathcal V$ denotes the participating nodes with $n = \abs{\vc}$, and $\mathcal E$ the set of edges, or payment channels, in the PCN. 
Each edge $(u,v)\in \mathcal E$ is associated with two weights, $b_{uv}$ and $b_{vu}$, which denote the balances from $u$ to $v$ and from $v$ to $u$, respectively. 
The \emph{capacity} of the channel, denoted as $C_{uv}=b_{uv}+b_{vu}$, is assumed to be a constant; this models a setting where the channel remains open for the entire duration of the experiment.
Recall that our goal is to release noisy channel balances, so the true balances may not be equal to the publicly-released channel balances, which we denote by $\tilde b_{uv}$ and $\tilde b_{vu}$, respectively.


We assume an arbitrary sequence of transactions $x_1, x_2, \ldots$ enters the system sequentially. 
Each  transaction $x_i$ has an associated source $s(x_i)$,  destination $d(x_i)$, amount $r(x_i)$, and timestamp $t(x_i$), where $t(x_i) < t(x_{i+1}) $ for all $i$.
Each transaction is processed instantaneously at its time-of-arrival timestamp. 
That is, we do not account for concurrent transactions. 
For a path $P$ on the graph, we use $s(P)$ and $d(P)$ to denote its source and destination, respectively.
At transaction $x$'s time of arrival, $t(x)$, the source $s(x)$ chooses a path $P$ from the network with source $s(P)=s(x)$ and destination $d(P) = d(x)$, such that $P$ appears to have enough balance according to the publicly visible balances. 
That is, $\forall (u,v)\in P$, $\tilde b_{uv} \geq r(x)$.
Recall that channel balances are noisy, so a path that appears to have enough balance may not in reality. 

If the path chosen for transaction $x$ does not have enough balance (i.e., $\exists (u,v) \in P: b_{uv}< r(x)$), the transaction fails. In this case, both the visible channel balances $\tilde b_{uv}$ and the true channel balances $b_{uv}$ remain unchanged. 
If the path chosen for transaction $x$ has enough balance (i.e., $\forall (u,v) \in P,  b_{uv}\geq r(x)$), the transaction succeeds, and $\forall (u,v)\in P$, the true channel balance is updated as $b_{uv} = b_{uv}-r(x)$ and $b_{vu}=b_{vu}+r(x)$.
The visible channel balances, on  the other hand, are updated according to a \emph{noise mechanism} (or \emph{mechanism}).
The noise mechanism is probabilistic and chosen by the system designer; 
given an input path $P$, it outputs a random set of edges $Q \subseteq P$ such that $\forall (u,v)\in Q$, the public balance $\tilde b_{uv}$ is updated to the true new balance, $b_{uv}$. 
We denote this conditional probability distribution by $\pol{P}{Q}$.

Notice two things about this model: first, we never update public balances on edges that were not involved in a transaction, i.e., edges that are not elements of path $P$.
This modeling choice was made in part for analytical tractability and in part for practicality; since nodes anyway must communicate with all relays in $P$, it is easy to include an instruction for the noisy balance update. 
Reaching other nodes would require additional communication that may not be practical. 
Second, note that if an edge is updated, it is only updated to its true balance---never a noisy version of its balance. 
One could additionally impose the condition that after choosing a subset $Q$, the noise mechanism updates the balances in $Q$ by different amounts (not necessarily equal to $r(x)$); however, as we will discuss in Section \ref{sec:model:prv}, our privacy metric only considers adversaries who aim to identify the source or destination of a transaction---not the amount.
If an adversary observes \emph{any} update on an edge, it knows that edge was involved in the transaction; knowing the exact amount does not give any more information about the source or destination.
Hence, to maximize utility, we might as well reveal the full balance update.



\subsection{Privacy Metric} \label{sec:model:prv}
Our adversary is an  honest-but-curious user that passively observes the network and tries to infer the source and destination of the first transaction $x$ to pass through the system; 
the adversary does \emph{not} try to guess the transaction amount $r(x)$.
\blue{We make this modeling choice for simplicity and because transaction patterns often matter  more than transaction amounts; for instance, the fact that one sends money to an abortion clinic or to a political organization can be more telling than the actual amount transferred.}
We assume that the initial public balances are all identical to the real balances, i.e., $\forall (u,v)\in \mathcal E$, $b_{uv}=\tilde b_{uv}$ at time 0, and all the balances in the PCN are sufficient to support $x$; 
hence, this metric is equivalent to considering an arbitrary transaction $x$ but giving the adversary knowledge of the true balances shortly before $t(x)$.
More precisely, once $x$ has been processed, the adversary guesses one node $v \in \mathcal V$ with probability $\A[v|Q]$, where $\A$ is a randomized adversarial strategy. 
If $ v\in \partial P$, where $\partial P \triangleq \{s(P),d(P)\}$, i.e., if the adversary guesses the source \emph{or} destination correctly, it wins.
To avoid making assumptions about the transaction or PCN distribution, we consider a worst-case metric over both. 
As shown in (\ref{eqn:prv}), privacy is defined as the minimax probability that the adversary makes a wrong estimate:
\begin{align}
    \prv(\D) &= 1 - \sup_{\A} \min_{P \in \pc} \sum_{v \in \partial P, Q \subseteq P} \pol{P}{Q} \adv{Q}{v} \notag \\
    & = 1 - \sup_{\A} \min_{P \in \pc} \sum_{Q \subseteq P} \pol{P}{Q} \adv{Q}{\partial P},
    \label{eqn:prv}
\end{align}
where $\mathcal P$ denotes the set of paths in $\mathcal G$. For convenience, we let $\A[\partial P|Q]$ denote $\sum_{v\in\partial P} \A[v|Q]$.


Notice that $\A[\cdot|Q]$ is a probability distribution with event space $\vc$, i.e. the supremum over $\A$ is taken in the space 
\[ \mathcal A = \brs{\A}{\forall u \in \vc, P \in \pc, Q \subset P: \A[u|Q] \ge 0,~~ \sum_{v\in \vc} \A[v|Q] = 1 }. \]

So constraint $\A \in \mathcal A$ is an intersection of linear constraints. In addition, the objective of \eqref{eqn:prv} is the minimum of finite number of affine functions of $\A$. Hence, the optimization problem in \eqref{eqn:prv} is equivalent to an LP. Because $\mathcal A$ is compact, the supremum of the optimization problem is attained. 
However, solving this optimization is intractable in general, due to the optimization over all the paths in the PCN.  
We will show that under certain restrictions, a solution can be computed efficiently.








\subsection{Utility Metric} \label{sec:model:uty}



The performance of a PCN is commonly measured by its \emph{success rate}, or the fraction of transactions it successfully processes \cite{malavolta2017silentwhispers,roos2017settling,sivaraman2018routing}. 
However, expected success rate is a complicated function that depends on transaction workloads, graph topology, and initial balances.
Since we wish to avoid making assumptions on these characteristics, we instead consider a simpler notion of utility: how representative are the observed balances of the true underlying balances?  
Specifically, given the true balance of a link, we consider the probability that the observed balance is equal to the true balance, which we will refer to as the \emph{truthful probability}. 
In general, this probability for a given PCN still depends on the transaction workload, as well as parameters like the path length.
Hence, we consider a \emph{worst-case} transaction workload that minimizes the truthful probability.



To this end, we define the quantity in \eqref{eqn:uty} below as the utility of mechanism $\D$. 
It equals the minimum probability, over all paths and edges in paths, of a public edge balance equaling the true edge balance after a transaction passes through it:
\begin{equation}
    \label{eqn:uty}
    \uty(\D) = \min_{P \in \pc} \min_{\epsilon \in P} \sum_{Q: Q \ni \epsilon, Q \subset P} \pol{P}{Q}
\end{equation}
In the proposition below, we show how this notion of utility is related to the truthful probability.
\begin{proposition}
    \label{thm:tptn}
    The truthful probability of any edge given any value of the true balance is lower bounded by the utility $\uty(\D)$.
\end{proposition}


\pf We focus on a particular edge $\epsilon \in \ec$. Assume \textit{at the current state}, the true and public balances of $\epsilon$ are $B_\epsilon,~ \tilde B_\epsilon$, respectively. Define event
\begin{align*}
    E_\epsilon \triangleq~ & \text{there exists a previous transaction going through $\epsilon$.}
\end{align*}

   If $\neg E_\epsilon$ happens, then $\P[B_\epsilon = \tilde B_\epsilon| \neg E_\epsilon] = 1$, since both balances are identical to their initial states, at which time all public balances are truthful.
    
   If $E_\epsilon$ happens, we focus on the state when the last transaction $T_{-1}$ on path $P'$ involving $\epsilon$ was about to be launched. At this state, we assume the true and public balances of $\epsilon$ are $b_\epsilon$ and $\tilde b_\epsilon$, while they have been $B_\epsilon$ and $\tilde B_\epsilon$ since $T_{-1}$ was launched, respectively. By definition of utility,
    \[ \uty(\D) \le \sum_{Q: Q \ni \epsilon, Q \subset P'} \pol{P'}{Q} = \P[\tilde B_\epsilon = B_\epsilon|E_\epsilon, T_{-1}]. \]
    Since $T_{-1}$ could be arbitrary, $\P[\tilde B_\epsilon = B_\epsilon|E_\epsilon] \ge \uty(\D)$.

To summarize, $\P[B_\epsilon = \tilde B_\epsilon] \ge \uty(\D)$ holds for all $\epsilon \in \ec$ regardless of previous workload on the network. \qedb

From the proof of Proposition~\ref{thm:tptn}, we can also see that $\uty(\D)$ is \emph{equal} to the truthful probability of an edge $\epsilon$ in the following scenario: a transaction happens along a path $P'$ that contains $\epsilon$, and $\epsilon$ and $P'$ are the minimizers in the definition of $\uty(\D)$.  Therefore, $\uty(\D)$ characterizes the truthful probability in a worst-case sense. 
For this reason, we will focus on $\uty(\D)$ as our utility metric.

\blue{
A natural question is how this utility metric relates to the success rate of a PCN.
The answer is complicated, depending on a number of factors.
However,  we show in Section \ref{sec:relation} that  under certain workloads and network conditions, our utility metric and the success rate of a PCN appear to be monotonically related. 
Hence, maximizing utility is as good as maximizing success rate.
}

%% file: sections/3_tradeoff.tex
\section{Fundamental Limits} \label{sec:fundlimits}



As system designers, we want the the privacy and utility metrics defined in \eqref{eqn:prv} and \eqref{eqn:uty} to be high, i.e., close to 1.
In this section, we show that for an important class of routing algorithms, this is not possible; additionally, we provide a tight upper bound on the tradeoff between these quantities for general graph  topologies. 

Let us begin by considering two extreme points. 
First, consider a setting with perfect utility, $\uty(\D) = 1$. 
To achieve perfect utility, on every path, the balance of every edge must be truthfully updated.
Hence, if the adversary simply guesses the source of the observed updated path, it always wins, so we have no privacy: $\prv(\D)=0$.

Next, consider a case with no utility, $\uty(\D) = 0$. 
This implies that for every path, noise mechanism $\D$ always hides the balance update on at least one edge from the true path. 
Regardless of the noise mechanism, the adversary can always pick a node uniformly at random.  
Since there are $n$ nodes, it guesses the source or destination with probability $2/n$, so $\prv(\D) \le 1- \frac{2}{n},~~ \forall \D.$ 
Hence, we have an upper bound on the privacy metric at the two extremes of the utility spectrum.
 
\begin{figure}
    \centering
    \includegraphics[width=\mywidth]{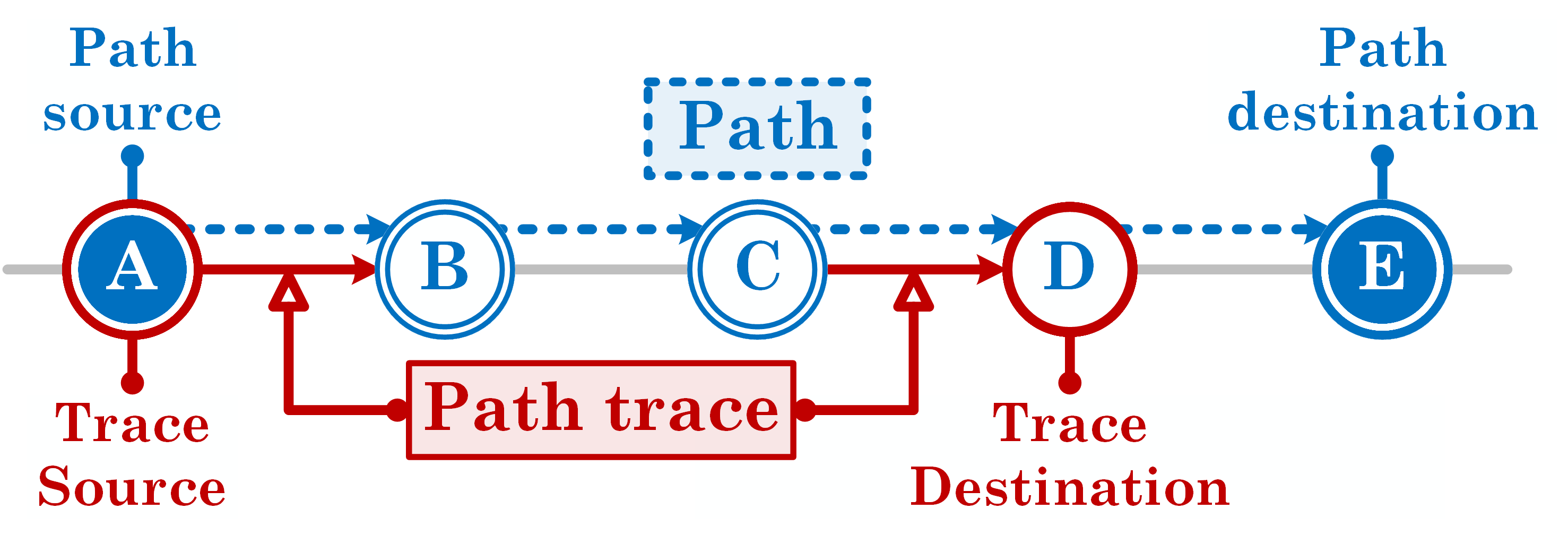}
    \caption{Example of a path from $A$ to $E$ and a corresponding path trace,
    the set of edges $A\to B$ and $C\to D$.}
    \label{fig:path_trace}
\end{figure}

The more challenging and interesting case arises when $0 < \uty(\D) < 1$.
We first define a \emph{path trace}, which intuitively describes a (possibly disjoint) set of edges that are elements of a path between two endpoints.
This set of edges must include the first and last hop of the path, adjacent to the two endpoints (Figure \ref{fig:path_trace}). 
\begin{definition}
    \label{def:isp}
    Given two nodes $x$ and $y$ on an arbitrary undirected graph $(\vc, \ec)$ with set of available paths $\pc$, a set of oriented edges $Q$ is a \textbf{path trace }from $x$ to $y$, if and only if 
    \begin{enumerate}
        \item $x$ has an incident edge in $Q$ with $x$ being the source;
        \item $y$ has an incident edge in $Q$ with $y$ being the destination;
        \item there exists a path $P \in \pc, P \supseteq Q$ from $x$ to $y$.
    \end{enumerate}
    
    We say $x$ is a \textbf{source} of $Q$, and $y$ is a \textbf{destination} of $Q$.
\end{definition}
In other words, a path trace is what the adversary sees after the noise mechanism updates a subset of edges in a given path. 
A path trace is always a subset of an available path in $\pc$. 
Our first main result, Theorem \ref{thm:linear}, states that if $\pc$ includes only shortest paths between nodes, then the privacy metric is upper bounded by a linear relation that we call the \emph{diagonal bound}.
Notice that shortest-path routing is extremely common.
First, today's PCNs like the Lightning network implement shortest-path routing by default \cite{lnd}.
Second, since intermediate relays in PCNs extract transaction fees, shortest-path routing can be thought of as a proxy for cheapest routing, which users are incentivized to use. 
Hence this result applies to the vast majority of routes used in practice. 

\begin{theorem}[The Diagonal Bound]
    \label{thm:linear}
    On a network $(\vc, \ec, \pc)$ with $n = \abs{\vc} \ge 2$, if $\pc$ includes only shortest paths, then for any noise mechanism $\D$, its privacy $\prv(\D)$ and utility $\uty(\D)$ satisfy
    \begin{equation}
        \label{eqn:diag}
        \prv(\D) \le 1 - \uty(\D). 
    \end{equation}
\end{theorem}

\begin{figure}[t]
    \centering
    \includegraphics[width=.6\mywidth]{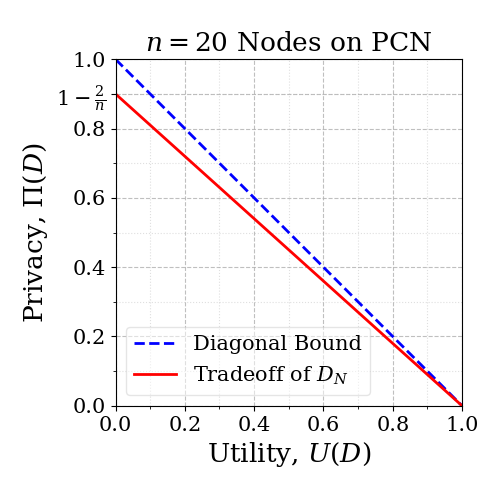}
    \caption{The diagonal upper bound of Theorem  \ref{thm:linear}, and the achievable tradeoff for the all-or-nothing scheme.
    \revisedel{The all-or-nothing curve is slightly  displaced from the upper bound for greater visibility.}} 
    \label{fig:lb}
    \Description{The Linear Bound.}
\end{figure}

\begin{figure}[b]
    \centering
    \includegraphics[width=\mywidth]{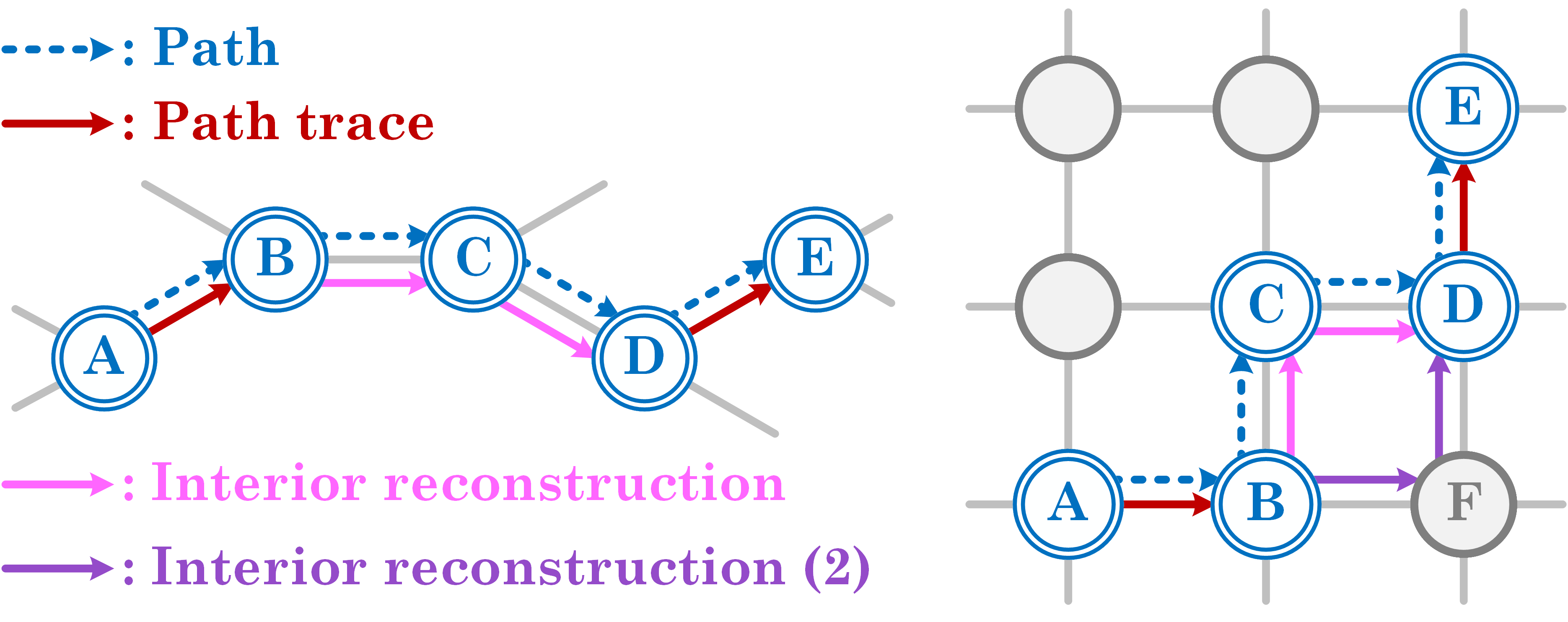}
    \caption{Examples of a path and a path trace on a 3-regular tree (left) and a grid graph (right), assuming shortest-path routing.
    For a tree, given any path trace, we can always uniquely determine the path that generated the path trace (up to the endpoints of the path trace).
    On a grid graph (indeed, on general graphs), we can uniquely determine the endpoints of the path trace, but not necessarily the full path.}
    \label{fig:fbex}
\end{figure}

Notice that this bound, plotted in Figure \ref{fig:lb}, does not make assumptions about the structure of the underlying graph. 
Intuitively, the result holds because for any observed path trace taken from a shortest path between two nodes on an arbitrary graph, the adversary can uniquely identify its source and destination.
For example, consider Figure \ref{fig:fbex} (left). 
On a tree, even if the noise mechanism reveals only a subset of edges in a path, the adversary can trivially reconstruct the interior path exactly. 
From this reconstruction, the endpoints are clear. 
Notice that in a path trace, we know the orientation of each edge because a truthfully-updated channel will always reveal the direction of money flow.
In Figure \ref{fig:fbex} (right), we see that on a grid graph, the adversary cannot always reconstruct the exact path, as there may be multiple shortest paths consistent with the path trace. 
For instance, the adversary cannot tell if $B\to C \to E$ or $B \to D \to E$ was the true path.
    However, the adversary \emph{can} uniquely determine the set of endpoints of the trace: $A$ and $E$. 
    Any other choice (e.g, $A$ and $D$) that passes through the full set of path trace edges has a strictly longer path length.
    This observation is true for general graphs, allowing the adversary to use the source or destination of the path trace as its estimate of the source or destination of the original path. 
This in turn causes the privacy metric for a noise mechanism to be governed primarily by its behavior near the source and destination, giving rise to the diagonal bound. 

We next prove that the bound in Theorem \ref{thm:linear} is tight \revise{as $n \to \infty$}. 
The proof of the upper bound implies that hiding edges in the interior of a path does not improve privacy if one or both of the endpoints of the path are revealed.
In other words, an adversary cannot always determine the sequence of edges between the endpoints of a path trace, but it can identify the \emph{endpoints} of the trace. 
Because there is no privacy benefit to hiding interior edges of a path, we should maximize utility (for a given privacy level) by revealing all interior edges. 
This motivates the so-called \emph{all-or-nothing} noise mechanism.
\begin{definition}
    \label{def:aon}
    For a transaction routed over path $P$, the \textbf{all-or-nothing noise mechanism} $\D\nve$ either truthfully updates balance on every edge of $P$ with probability $\uty(\D\nve)$, or updates nothing with probability $1-\uty(\D\nve)$. 
\end{definition}

We define a \emph{reachable} network as follows:
\begin{definition}
    \label{def:nfr}
    A network $(\vc, \ec, \pc)$ is \textbf{reachable}, if for any pair of nodes $(u, v) \in \vc^2$ with $u \neq v$, there exists a path $P \in \pc$ that goes from $u$ to $v$, or from $v$ to $u$. 
\end{definition}
\blue{Note that reachable networks are more restrictive than the notion from the literature of \emph{strongly connected networks}, because we limit the set of usable paths to $\pc$.}


\begin{lemma}
    \label{thm:unifguess}
    On a reachable network, 
    \begin{equation}
        \sup_{\A[\cdot|\emptyset]} \min_{P \in \pc} \A[\partial P | \emptyset] = \frac{2}{n}. \label{eqn:unifguess}
    \end{equation}
\end{lemma}

\pf Because $\A[\cdot|\emptyset]$ is a probability distribution, for any strategy $\A$, there exist 2 nodes $u_1, u_2$ such that
\[ \A[u_1|\emptyset] + \A[u_2|\emptyset] \le \frac{2}{n}. \]

By assumption that the network is reachable, there exists a path $P_u$ connecting $u_1, u_2$. Hence,
\[ \min_{P \in \pc} \A[\partial P|\emptyset] \le \A[\partial P_u|\emptyset] \le \frac{2}{n},~~ \forall \A. \]

In the meantime, the adversary who guesses each node uniformly could make the inequality above tight, which proves \eqref{eqn:unifguess}. \qedb

On a reachable network, the below theorem characterizes the privacy-utility tradeoff for the all-or-nothing noise mechanism. 

\begin{theorem}
    \label{thm:dntradeoff}
    On a reachable network, the privacy-utility tradeoff of the all-or-nothing noise mechanism is characterized by
    \begin{equation}
        \label{eqn:dntradeoff}
        \prv(\D\nve) = \bra{1 - \frac{2}{n}} [1 - \uty(\D\nve)]. 
    \end{equation}
\end{theorem}
Although this result holds for all reachable networks, regardless of routing policy, 
it also implies that the upper bound in Theorem \ref{thm:linear}, 
\revise{$\prv(\D) \le 1 - \uty(\D)$,} 
is \revise{asymptotically} tight for shortest-path routing over reachable networks. 

\pf When the entire path is updated, the adversary $\A_0$ may directly pick an endpoint of the observed path as its estimate. 
Hence,
\[ \sup_{\A} \min_{P'\in\pc} \A[\partial P'|P'] = \A_0[\partial P|P] = 1, \quad \forall P \in \pc. \]

By our privacy definition,
\begin{align}
    \prv(\D\nve) & = 1 - \sup_{\A} \min_{P \in \pc} \sum_{Q \subset P} \D\nve[Q|P] \A[\partial P|Q] \notag \\
    & = 1 - \sup_{\A} \min_{P \in \pc} \brc{ \brb{1-\uty(\D)} \A[\partial P|\emptyset] + \uty(\D) \A[\partial P|P] } \notag \\
    & = [1 - U(\D)]\bra{1 - \sup_{\A} \min_{P \in \pc} \A[\partial P|\emptyset]} \notag \\
    & = [1 - U(\D)]\bra{1 - \frac{2}{n}}. \tag{*} \label{eqn:srtlimitle} \notag
\end{align}




Obtained by Lemma \ref{thm:unifguess}, \eqref{eqn:srtlimitle} justifies the original claim. \qedb

The results of this section are not encouraging; they imply that a perfect privacy-utility tradeoff is impossible as long as nodes use shortest-path routing. 
Since there is a monetary cost associated with taking longer paths, most users are likely to default to shortest-path routing. 
Nonetheless, some users do care about privacy and may be willing to pay for it \cite{beresford2012unwillingness,acquisti2013privacy}. 
The next section explores various techniques for breaking the diagonal bound, and the costs associated with doing so. 

%% file: sections/4_clique.tex
\section{Breaking the diagonal barrier}
\label{sec:barrier}

The goal of this section  is to break the diagonal barrier imposed by  Theorem~\ref{thm:linear} 
by relaxing two key assumptions: one related to adding uncertainty to the candidate endpoint set, and the other related to adding  uncertainty to the set of candidate paths. 
More precisely, Theorem~\ref{thm:linear} makes two main assumptions:
 $(i)$ utility $\uty(\D)$ is computed as a minimum over all network edges, as in the definition~\eqref{eqn:uty}, and 
$(ii)$ all transactions are routed using a shortest path between the source and destination. 
We will relax the first assumption to introduce endpoint uncertainty, and the second to introduce path uncertainty.

\emph{Assumption (i):} Definition \eqref{eqn:uty} of utility may not be appropriate when the PCN includes endpoint nodes with degree one, or clients that connect to ``gateway" routers.  
This could happen, for example, if PCNs evolve so that merchants run the majority of router nodes, which maintain several well-funded channels, and end users connect to merchants only for their own transactions, i.e., without participating in transaction relaying. 
As an extreme example, consider a star network, which is a tree of depth one. 
Any routing of a transaction requires at most two edges, whose balances are already known to the 
nodes directly involved in the transaction, i.e., the sender and the receiver. 
Because the sender and receiver already know the balances on their adjacent edges, they could transmit this information  out of band, so hiding the balance of any edge does not decrease success rate, but provides perfect privacy against an external observer. 
In Section~\ref{sec:usm}, we show how to improve the  privacy-utility tradeoff by allowing the source and destination to exchange their adjacent true balances prior to transacting. 

\emph{Assumption (ii):} The second assumption might not hold if users 
are willing to route  transactions over paths longer than the shortest path(s).  
At the cost of incurring more fees, such longer routing 
can intuitively achieve better privacy-utility tradeoff. 
Although the gains are difficult to quantify in general, we precisely quantify them for a special network topology in Section~\ref{sec:clique}.


\newcommand{\usm}{User-Server Model}
\subsection{Endpoint Uncertainty}
\label{sec:usm}

In real-world PCNs, users know the true balances (or weights) on their directly adjacent channels (or edges). 
For example, when a user Alice tries to transact with neighboring user Bob, Alice and Bob already know each other's true balances. 
Therefore, for any direct transaction among neighbors, success rate is not sacrificed even if the balances are kept private. 
In an extension of this idea, consider a PCN model with \emph{users} and \emph{servers}; 
users make transactions among themselves, whereas servers relay them. 
Servers are connected in a network structure $(\vc_S, \ec_S)$,   
whereas each user node is connected to one server node. 
Let $\vc_U, \ec_U$ denote the set of user nodes and user-server channels, respectively. 
Then there exist functions $J:\vc_U \to \vc_S$ and $K:\vc_U \to \ec_U$, where for a user node $v\in \vc_U$, $J(v)$ is the server to which $v$ is connected, and $K(v)$ is its unique incident channel, which connects $v$ and $J(v)$. 
Furthermore, let $(n_S, n_U) = (\abs{\vc_S}, \abs{\vc_U})$. 
The public balances on the user-server channels are never updated, while those on the inter-server channels are updated with a privatizing noise mechanism of our choice.

Under the following assumption that the set $\pc$  of available paths are not too restricted, we show that just privatizing the inter-server channels is sufficient to achieve improved privacy-utility tradeoff.  

\newtheorem{propo}{Proposition}[]
\newtheorem{assmp}[propo]{Assumption}

\begin{assmp} 
 \label{assmp:usm} 
For any oriented path $P \in \pc$ from $x$ to $y$, 
    \begin{itemize}
        \item if $x \notin \vc_S$, then $P - \brc{(x, J(x))} \in \pc$;
        \item if $x \in \vc_S$, then $\forall v : J(v) = x$, we have $\brc{(v, x)} \cup P \in \pc)$;
        \item if $y \notin \vc_S$, then $P - \brc{(J(y), y)} \in \pc$;
        \item if $y \in \vc_S$, then $\forall v : J(v) = y$, we have $\brc{(y, v)} \cup P \in \pc)$.
    \end{itemize}
\end{assmp}
\blue{Assumption \ref{assmp:usm} states that if an oriented path from server $x'$ to server $y'$ is observed, the adversary cannot obtain more information from $\pc$ about the source or destination. 
For example, he only learns that the source can be server $x'$ itself or one of its incident users. }

\begin{proposition}
    \label{thm:usm}
    Suppose we have a user-server network $(\vc_S \cup \vc_U, \ec_S \cup \ec_U)$, and the set $\pc$ of available paths satisfy  Assumption~\ref{assmp:usm}, 
     the server sub-network $(\vc_S,\ec_S)$ is reachable under $\pc$, 
    and all the transactions are routed via 
     the shortest paths. 
    When the channels on the server network $(\vc_S, \ec_S)$ 
    have balances governed by the all-or-nothing noise mechanism $\D\nve$ with utility $\uty(\D\nve) = \alpha$, 
    and the channels on the user network $(\vc_U,\ec_U)$ are completely hidden, the privacy metric  $\prv(\D\nve)$ is  
    \begin{equation}
        \label{eqn:usm}
        \prv(\D\nve) = \bra{1 - \frac{2}{n_S + n_U}} \bra{1 - \alpha} + \frac{\uu}{\uu+1} \alpha \;,
    \end{equation}
    where $\uu$ is a lower bound on the number of users attached to any server node. 
    
\end{proposition}

Compared with all-or-nothing noise on a full PCN (Theorem \ref{thm:dntradeoff}), the privacy level of the user-server model is higher by $\mu \uty(\D) / (1+\mu)$. 
Figure~\ref{fig:perf}, left panel,  plots  the gains achievable from Theorem \ref{thm:usm} under all-or-nothing noise.
As the number of users per server increases, we get improved tradeoffs, eventually achieving maximum privacy with no utility loss in the limit as $\mu, n_S, n_U\to\infty$. 
Intuitively, the adversary cannot distinguish between sources or destinations connected to the first and last server nodes. Since the adjacent user nodes as well as the last server node can all be the real destination of money flow, the adversary's best strategy is to guess uniformly at random. 
Hence the gain in privacy comes from the increased uncertainty at the end nodes, which does not sacrifice the success rate of routing transactions. 
This observation can be generalized to each user having multiple connections to servers, as long as the users do not relay transactions. 
Moreover, preliminary evidence suggests that user-server models may naturally emerge in practice; thus far, we have seen the emergence of routing hubs on the Lightning network, complemented by many users with few channels\apdx{ (Figure \ref{fig:lnd})}. 

\begin{figure}[ht]
    \centering
    \includegraphics[width=.7\mywidth]{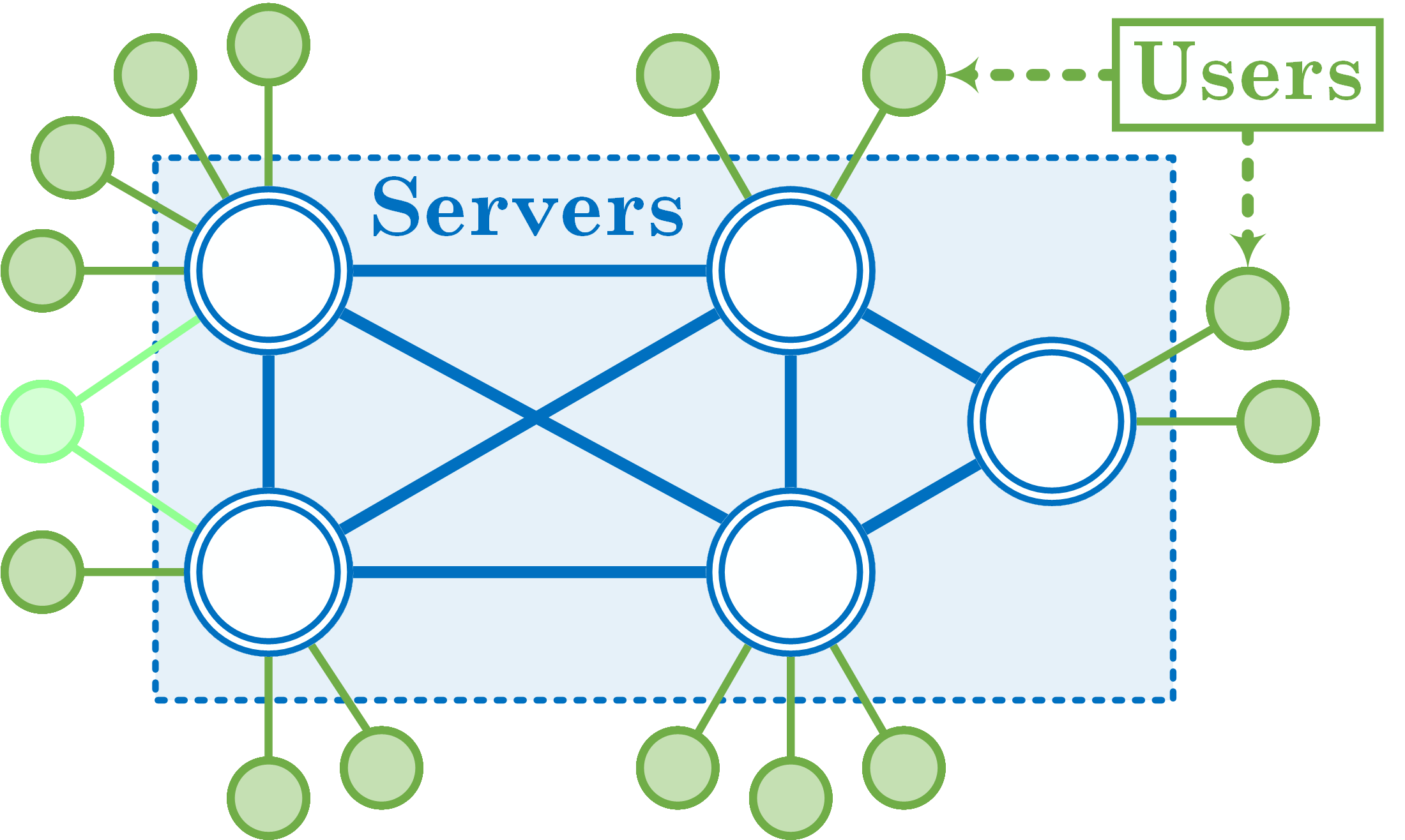}
    \caption{\blue{The user-server model consists of server nodes that route transactions, and user nodes who make transactions, but do not relay other nodes' transactions.}}
    \label{fig:usm}
\end{figure}

\blue{
Proposition \ref{thm:usm} holds when each user is allowed to be connected to only one server. 
In this case, the size of server sub-network is limited, since $n_S \le n / (\mu +1)$. 
We can easily extend this result to the case where each user node is allowed to share channels with multiple server nodes, but users do not relay transactions.
\wt{For example, in Figure \ref{fig:usm}, the light-green user is allowed to maintain two channels with two different servers. }
In this case, the network structure will be much more compact, and its privacy is only slightly worse than \eqref{eqn:usm}, as seen in the following:

\begin{proposition}
    \label{thm:usmv2}
    Assume the assumptions of Proposition \ref{thm:usm} except that each user may connect to multiple servers (but not relay transactions). 
    The privacy metric $\prv(\D\nve)$ of the all-or-nothing noise $\D\nve$ satisfies
    \begin{equation}
        \label{eqn:usmv2}
        \prv(\D\nve) \ge \bra{1 - \frac{2}{n_S + n_U}} \bra{1 - \alpha} + \frac{\uu}{\uu+2} \alpha \;,
    \end{equation}
    where $\uu$ is a lower bound on the number of user nodes attached to any server node. 
\end{proposition}
}
This suggests that in practice, user endpoints can improve their privacy by not updating their balances without losing any utility, as long as they have an out-of-band channel of communication with the recipient of a transaction.
This may not always be feasible, e.g.,  in cases where the sender or recipient wishes to remain anonymous.

\subsection{Path Uncertainty} \label{sec:clique}
The previous subsection showed how a special class of networks can be exploited to increase user privacy by introducing uncertainty regarding the endpoints of a path under all-or-nothing noise. 
In this section, we explore techniques for obfuscating the transaction path itself. 
When users are allowed to use longer paths,  
the noise mechanism can hide edges so that path traces do not reveal the source/destination pair. 
This is not true under the shortest path routing as illustrated in Section~\ref{sec:fundlimits}. 
For the rest of this section, we consider the special case of a \emph{complete graph}. 
Although complete graphs are not realistic for deployed PCNs, 
they provide a method for analyzing the privacy definition LP in \eqref{eqn:prv}, as well as inspiration for noise mechanisms that can break the diagonal barrier in Sec. \ref{sec:fundlimits}. 

\begin{definition}
    Given the edges on a directed path $P$, let the \emph{odd} edges include the 1st, 3rd, 5th, ... edges and the \emph{even} group include the 2nd, 4th, 6th, ... edges. The \textbf{alternating noise mechanism} $\D\alt$ is defined as follows:
    \begin{itemize}[leftmargin=*]
        \item At utility  $\alpha \in [0, 0.5]$, $\D\alt$ either updates the balances of the odd edges with probability $\alpha$, updates balances of the even edges with same probability, or updates nothing with probability $1-2\alpha$. 
        \item At utility  $\alpha \in (0.5, 1]$, $\D\alt$ either updates the balances of the odd edges with probability $1-\alpha$, updates balances of the even edges with same probability, or updates nothing with probability $2\alpha-1$. 
    \end{itemize} 
\end{definition}

\begin{figure}
    \centering
    \includegraphics[width=.4\mywidth]{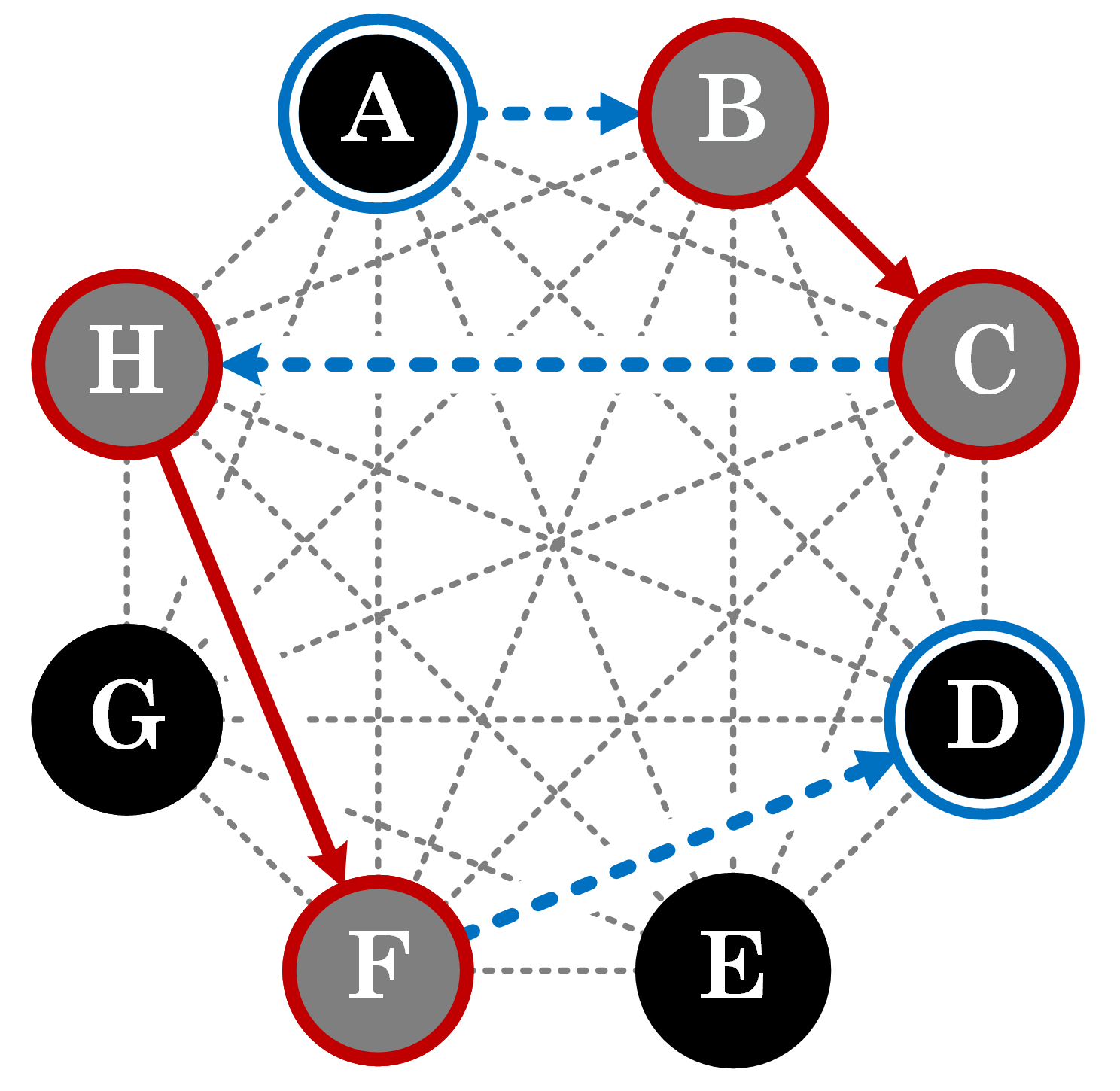}
    \caption{A path and path trace for the alternating noise mechanism; in this case, even edges were revealed.}
    \label{fig:alt}
\end{figure}

To preserve symmetry, $\pc$ is assumed to be the set of all simple paths of fixed length $L$. 
Note that it may be computationally infeasible to find a simple path of given length that satisfies a balance constraint in practice. 
Intuitively,  we expect alternating noise to have good privacy;
even when the adversary knows the path length $L$,  each revealed edge could be the first or last edge, and the graph topology gives no way to differentiate between edges. 
For example, in Figure \ref{fig:alt}, the adversary is able to discern that the even edges were revealed since $L=5$ and only two edges are revealed. 
Hence it must guess the source/destination uniformly from the nodes that are not part of the path trace. 
If the odd edges had been revealed, then because the graph is complete, any ordering of the edges would have been feasible.
Hence, the adversary would need to guess one of the endpoints of the revealed edges uniformly at random.
Combined, these cases give asymptotically perfect privacy as $L,n \to \infty$. 
The next theorem characterizes this tradeoff precisely.
\begin{theorem}
    \label{thm:alttradeoff}
    Suppose $L \ge 2$ and the alternating scheme $\D\alt$ has utility $\alpha = \uty(\D\alt)$ and privacy $\prv(\D\alt)$. If $2 \mid L$, i.e. $L$ is even, its privacy-utility tradeoff on a complete graph  is characterized by
    \begin{equation}
        \prv(\D\alt) = \left\{ \begin{array}{lr}
            1-\dfrac{2}{n} - \bra{\dfrac{2}{\min\brc{L, n-L}} - \dfrac{4}{n}}\alpha, & 0 \le \alpha \le \dfrac{1}{2}; \\
            \bra{2-\dfrac{2}{\min\brc{L, n-L}}}(1-\alpha), & \dfrac{1}{2} < \alpha \le 1.
        \end{array} \right. 
        \label{eqn:alteven}
    \end{equation}
    
    If $2 \nmid L$, i.e. $L$ is odd, its privacy-utility tradeoff is characterized by
    \begin{equation}
        \prv(\D\alt) = \left\{ \begin{array}{lr}
            1-\dfrac{2}{n} - \bra{\dfrac{2}{L+1} + \dfrac{2}{n-L+1} - \dfrac{4}{n}}\alpha, & 0 \le \alpha \le \dfrac{1}{2}; \\
            \bra{2-\dfrac{2}{L+1}-\dfrac{2}{n-L+1}}(1-\alpha), & \dfrac{1}{2} < \alpha \le 1.
        \end{array} \right. 
        \label{eqn:altodd}
    \end{equation}
\end{theorem}

\begin{figure}[t]
    \centering
    \includegraphics[width=\mywidth]{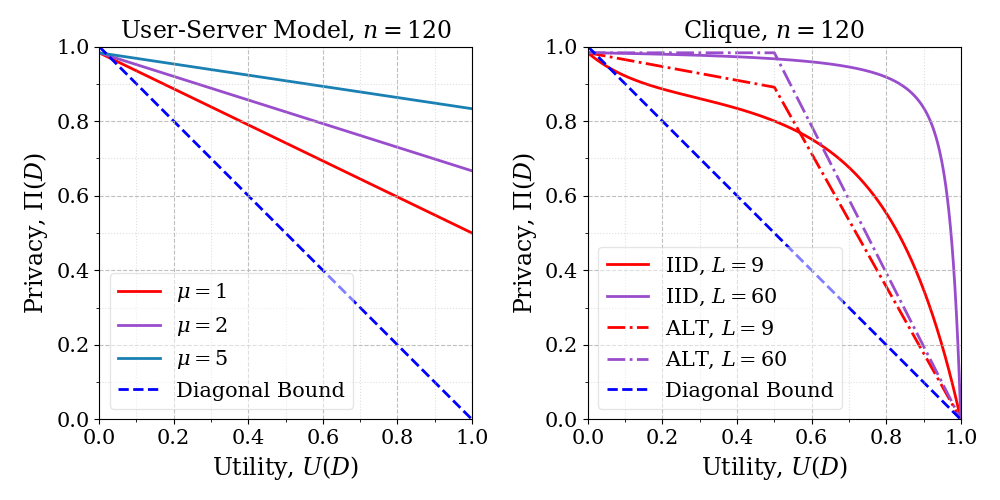}
    \caption{Exploiting the structures of the networks, we can design mechanisms that overcome the diagonal bound of Theorem~\ref{thm:linear} (shown in blue dashed line). Under the user-server model, we exploit endpoint uncertainty and under the clique model, we exploit path uncertainty. } 
    \label{fig:perf}
\end{figure}

This tradeoff is plotted in Figure \ref{fig:perf} (right) for two path lengths,  $L=9$ and $L=60$. 
Notice that the difference between these curves is small, even though the difference in path lengths is almost an order of magnitude. 
Hence, using longer paths appears to have diminishing returns for alternating noise. 

Another natural mechanism which requires less coordination among the nodes in a path is the \emph{i.i.d. noise mechanism}. 


\begin{definition}
    The \textbf{i.i.d. noise mechanism} $\D\iid$ updates the balance of each edge independently with constant probability $\uty(\D\iid)$. Mathematically, for a path $P$ with length $L$,
    \[ \D\iid[Q|P] = [\uty(\D\iid)]^{\abs{Q}} [1-\uty(\D\iid)]^{L-\abs{Q}},\quad \forall Q \subset P. \]
\end{definition}
This noise mechanism enjoys similar privacy gains as the alternating mechanism due to path uncertainty. 
The main difference is that 
the number of candidate endpoints is now a random variable. 
We characterize the utility/privacy tradeoff precisely below. 

\begin{theorem}
    \label{thm:iidc}
    Suppose $(\vc, \ec)$ is a complete graph and $\pc$ is the set of all simple paths of fixed length $L$. The privacy $\prv(\D\iid)$  of i.i.d. noise mechanism $\D\iid$ on a complete graph is lower bounded by
    \begin{equation}
    \label{eqn:clq}
        \prv(\D\iid) \ge 1 - \frac{2(1-\alpha)^{L}}{n} - \frac{2}{(L+1)(1-\alpha)} \brb{1-\alpha^{L+1} - (1-\alpha)^{L+1}} \;,
    \end{equation}
    where $\alpha= \uty(\D\iid)$ is the utility. 
\end{theorem}



Figure~\ref{fig:perf} (right) plots the achievable bounds of  Theorems  \ref{thm:alttradeoff} and \ref{thm:iidc}, illustrating that neither scheme strictly dominates the other. 
However, i.i.d. noise has a larger jump in performance as we move from paths of length $L=9$ to $L=60$.
\apdx{Due to space constraints, the proofs of Theorems \ref{thm:alttradeoff} and \ref{thm:iidc} can be found in Appendix \ref{app:proofs}.}

\paragraph{Takeaway Message}
In this section, we show that the diagonal bound can be broken through two techniques: adding endpoint uncertainty and adding path uncertainty. 
Our exploration of path uncertainty in particular is limited to complete graphs due to the computational complexity of evaluating privacy metric \eqref{eqn:prv} on general graphs. 
Still, we conjecture that similar intuitions can be applied to more complex, structured graphs; 
for instance, we relied on both the symmetry and the well-connected nature of complete graphs in our analysis; some circulant graphs exhibit similar properties. 
Hence, a takeaway message is that if one has the ability to choose a network structure, noise mechanism, \emph{and} routing policy, it is possible to achieve an improved  privacy-utility tradeoff;
\blue{although there is no central  entity controlling the graph topology, in  practice system designers can influence topology by implementing default peer selection rules in client software.}
However, these benefits come at the cost of higher 
transaction fees due to the choice of routing through longer paths.\footnote{\blue{In general, a shorter path can cost more than a longer path because routers choose their own routing fees. Our model ignores this distinction for simplicity and because on average, longer paths tend to cost more.}}
On the other hand, 
the user-server model arises organically  in current payment channels, and is compatible with cost-effective, shortest-path routing. 

%% file: sections/5_utility_metrics.tex
\section{Relation between our utility metric and success rate}
\label{sec:relation}
A natural question is how to relate these results to the more intuitive (but more complicated to compute) success rate metric in Section \ref{sec:model}.
In this section, we explore the relation between our utility metric and success rate and observe that under some conditions, they are positively correlated.

\begin{figure}
    \centering
    \includegraphics[width=.8\mywidth]{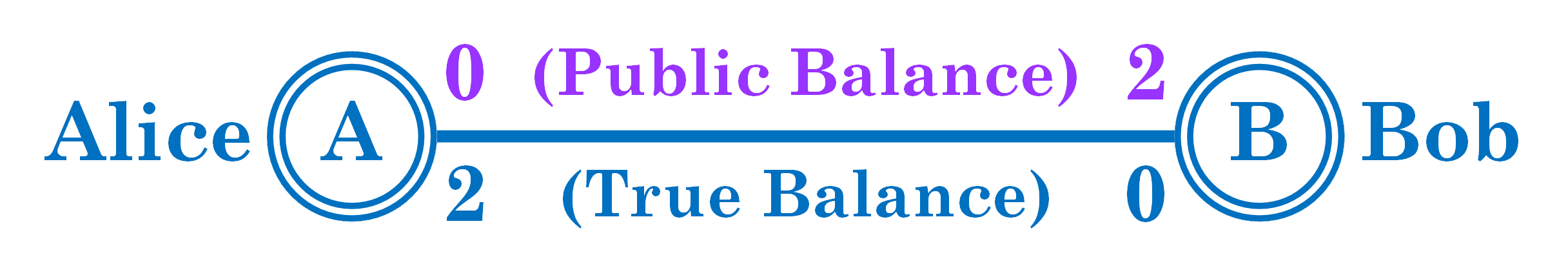}
    \caption{A deadlocked channel. No one else can send tokens through it in either direction. }
    \label{fig:dlkdemo}
\end{figure}

Upon initial study, these two metrics actually appear to be \emph{uncorrelated}
due to an interesting phenomenon we refer to as ``deadlocks.''  
A channel is deadlocked when the public and true balances concentrate at different ends of the channel such that it cannot support any more transactions. 
For example, as shown in Figure \ref{fig:dlkdemo}, let Alice and Bob maintain a channel with capacity $2$, where Alice holds $2$ tokens and Bob holds $0$. 
If the public balance shows the opposite, i.e., Bob claims to hold $2$ tokens and Alice $0$, 
routers will try to send money only from Bob to Alice; this is impossible because Bob actually has no balance. 
Since no transactions pass through the channel, it will remain deadlocked forever, reducing the success rate over this channel to zero.

\begin{figure}
    \centering
    \includegraphics[width=\mywidth]{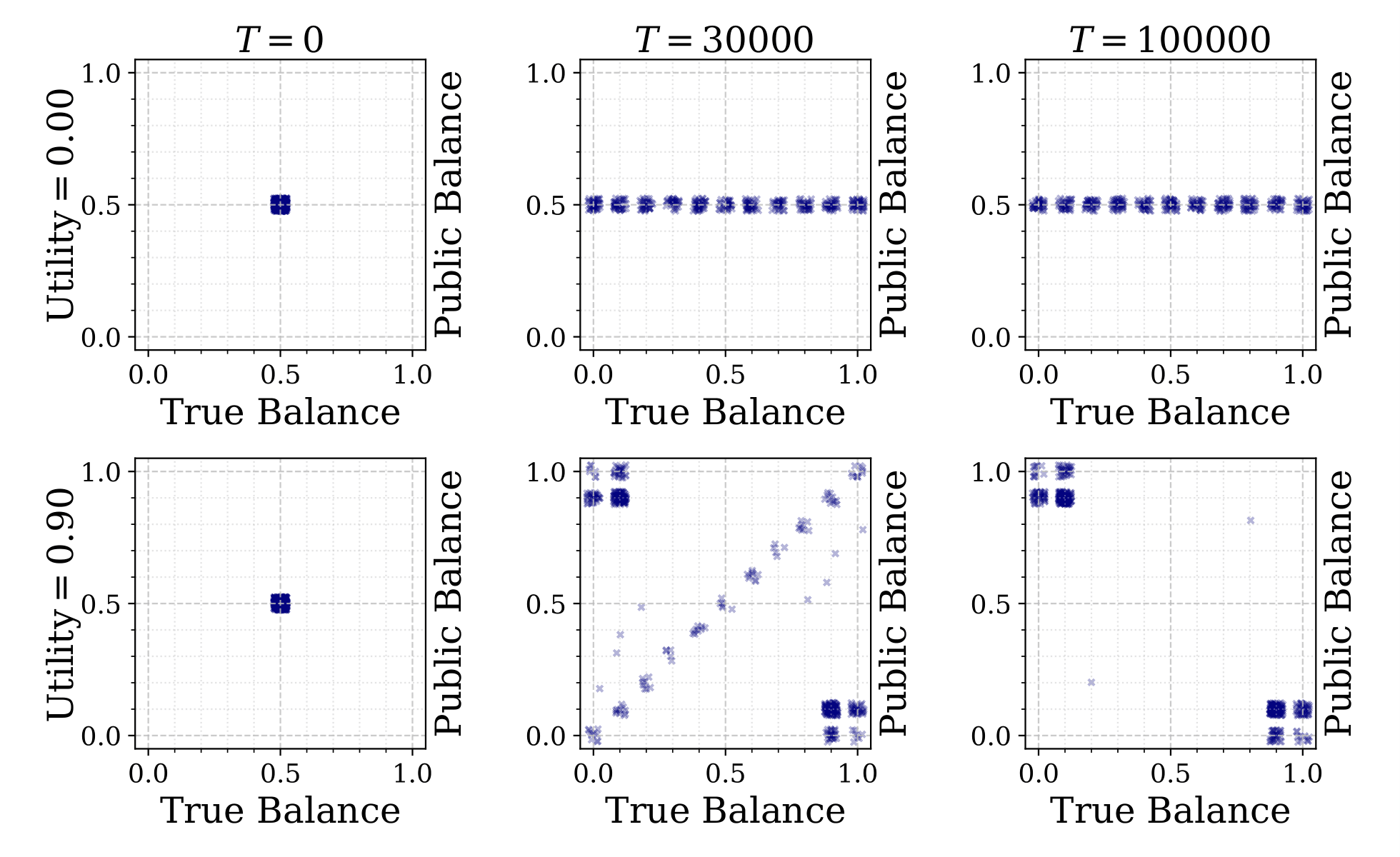}
    \caption{Scatter plot of true balance vs public balance on both ends of all channels in 200 parallel experiments. Each point is plotted with a small deviation to demonstrate the number of overlapping points.}
    \label{fig:dlkscatter}
\end{figure}

To demonstrate the deadlock effect, we show simulation results in the simplest setting, where the PCN consists of only $2$ nodes $A, B$ and $1$ channel. 
The channel has capacity $10$ and initial balance $5$ on both ends. 
Each transaction, denoted by $[\text{value}(\text{sender} \to \text{recipient})]$, is sampled from a uniform distribution over $\{2 (A\to B), 2 (B \to A), 3 (A \to B), 3 (B \to A)\}$.
The sender is assumed not to know the true channel balance. 
Figure~\ref{fig:dlkscatter} plots the true balances vs the observed ones for 200 such  independent channels as the number of transactions $T$ grows; 
a deadlocked channel will show balances at the top-left and the bottom-right corners. 
Note that even at a high utility $0.9$, almost all of the parallel single-edge PCNs are deadlocked after 100,000 transactions. 
In contrast, at perfect privacy (utility 0.0), no  deadlock occurs because public channel balances remain forever at their initial value, $(5, 5)$.

Unfortunately, naive noise implementations can lead to severely deadlocked networks. 
To formally quantify this effect, consider the following model: 
assume that each transaction (i.e., its sender, recipient, and value) is independently sampled from some probability distribution.  
We fix a specific routing scheme, e.g., randomized shortest path routing.
Under these assumptions, the process of transactions flowing through a PCN is a Markov chain. 
Each assignment of public and true balances is a state, and each transaction triggers a state transition. 
%
In the theorem below, we show that the steady-state success rate of this Markov chain is zero due to deadlocks.  
Intuitively, a sufficient number of channels will almost surely be eventually deadlocked after a finite number of transactions. 
These deadlocks block the entire PCN, resulting in a zero success rate.
\begin{theorem}[deadlock]
    \label{thm:zerosr}
    Suppose a PCN $\gc(\vc, \ec)$ is given with arbitrary topology and non-negative integer balances. 
    Each transaction is independently sampled from an arbitrary distribution with value at least $\ell \in \Z_{++}$.
    The sender and recipient are selected symmetrically, and neither of them knows the true balance of any channel. 
    Assume a noise mechanism $\D$ is applied to the transaction path, where the utility $\alpha \in (0, 1)$, and there exists an edge on the path, whose public balance is not updated with probability at least $\beta > 0$. 
    Then, if the number of transactions $T \to \infty$, the PCN's success rate is $0$.
\end{theorem}

\begin{figure}
    \centering
    \includegraphics[width=\linewidth]{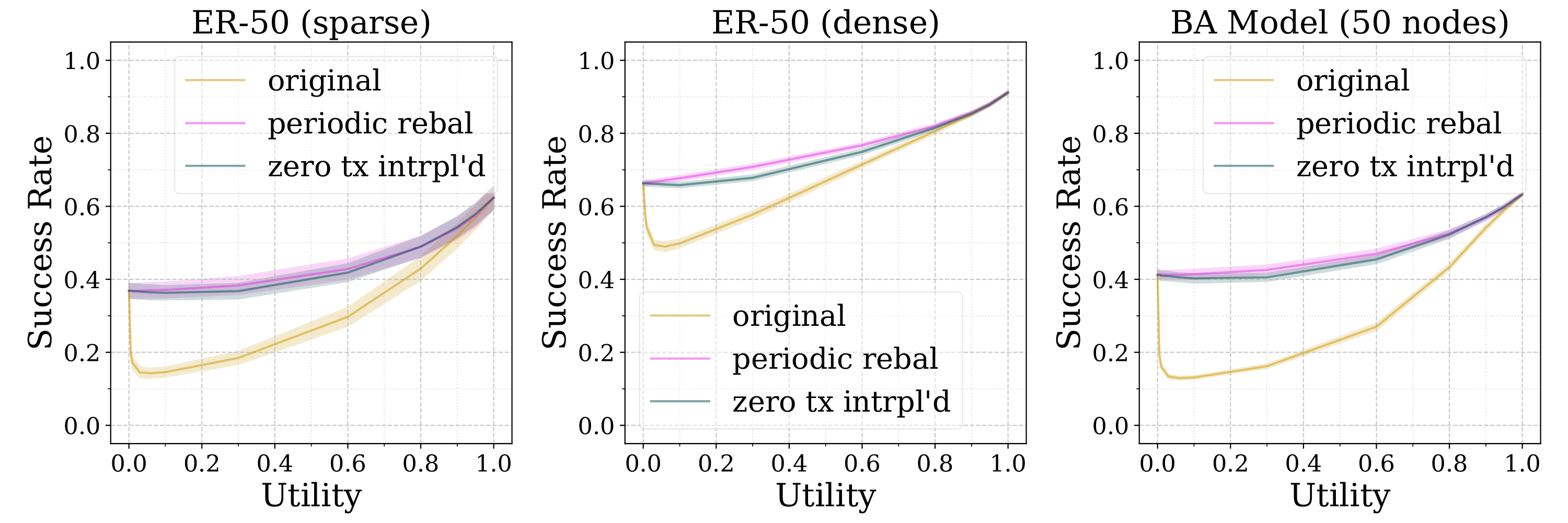}
    \caption{Relations between our utility metric and success rate with and without deadlocks. }
    \label{fig:alleviation}
\end{figure}

\apdx{(Proof in Appendix \ref{sec:pfzerosr})} 
Theorem \ref{thm:zerosr} holds because long-term,  \emph{every} channel will  become deadlocked and hence unusable. 
In particular, this implies that success rate is uncorrelated with utility. 
%
However, in practice, it is not true that route sources are unaware of balances of neighboring channels, or that there are infinitely many transactions.
Although the theorem no longer applies without these assumptions, its insight regarding deadlocks helps to explain the counterintuitive non-monotonicity of the utility-success rate curves in Figures \ref{fig:alleviation} and \ref{fig:exptop}: one would intuitively expect success rate to increase when routers have access to more information. 
When utility is close but not equal to $0$, the first edges on routes are likely to turn into deadlocks, negatively impacting success rate. 
If senders are aware of balances of neighboring channels, deadlocked channels can become unlocked by being the first edge on a route. 
This prevents success rates from falling to $0$, but it is too small an effect to completely cancel the influence of deadlocks. 

We consider two heuristic methods for alleviating deadlocks:

\noindent    \emph{(1) Periodic Rebalancing.} After every $m$ global transactions (i.e., counted across the whole PCN), each deadlocked channel resets its \textit{public} balances evenly to $(\frac{C}{2}, \frac{C}{2})$, where $C$ denotes channel capacity, with probability $0.5$. Otherwise, it remains deadlocked.

\noindent   \emph{(2) Zero-valued Transactions.} Introduce a new random process (e.g., Poisson) of $0$-valued transactions, whose endpoints are selected uniformly from the participating nodes. 
    These 0-transactions are subject to the same noise mechanism as regular transactions, and therefore update the public balances with some probability. 
    If the public balance on an edge is not updated to the true balance, that edge's balance is instead reset to $(\frac{C}{2}, \frac{C}{2})$.


\smallskip
\noindent Notice that neither of these heuristics affects the privacy guarantee or the utility of the noise mechanism.  
Our privacy metric assumes a worst-case setting, where the initial public balances before a transaction are the real balances, 
and our utility metric considers the public balance updates from a single transaction.
Hence, all  the results from  Sections \ref{sec:fundlimits} and \ref{sec:barrier} still apply. 
%
Nonetheless, we are not suggesting these countermeasures are ready to be put to use. 
For example, Periodic Rebalancing is difficult to deploy because in a decentralized distributed system, the nodes cannot easily know how many transactions happened globally. 
But most importantly, both heuristics leak balance information, introducing privacy drawbacks not captured by our metric. 
We leave analysis of these drawbacks to future work. 

Now, we wish to understand the effect of these heuristics on success rate for a given utility and noise mechanism. 
Characterizing this theoretically is challenging, even for simple graphs like complete graphs; as such, we turn to simulations, shown in Figure~\ref{fig:alleviation}.
Here, utility represents the utility metric of all-or-nothing noise mechanism applied to the PCNs, which is defined in Section~\ref{sec:fundlimits}. 
Success rate is estimated by the portion of successful transactions over the total number, 100,000. 
At each utility point, these transactions are run on 50 parallel PCNs. 
Each channel is initialized with 1 token on both sides, and each transaction is of value 1.
\wt{We test 3 different random topologies: 1) \ert{} networks with $50$ nodes and expected density $\log 50 / 50$; 2) \ert{} networks with $50$ nodes and expected density $0.25$; and 3) preferential attachment networks (Barab{\'a}si-Albert model). Initialized as a triangle, it has 47 new nodes inserted sequentially. Upon each insertion, the new node is attached to 2 existing nodes with probability  proportional to the degree of existing node. For simplicity, we denote the graph as $\ba(K_3, 47, 2)$, where $K_n$ denotes clique with $n$ nodes. }
Note that in these simulations, even without alleviating deadlocks, the success rate is not exactly zero.  
This is partially because our simulations are limited in duration, and partially because nodes are assumed to know the balances on their adjacent channels. 
Both of these properties violate the assumptions in Theorem~\ref{thm:zerosr},  but more realistically model a real PCN. 
%

%

The solid curves and their error bars represent the means and the standard deviations of estimated success rates over 50 parallel PCNs, respectively. 
The length of each error bar equals twice the corresponding standard deviation. 
Figure~\ref{fig:alleviation} shows that both methods successfully alleviate the effects of deadlocks, eliminating the drop in success rate due to deadlocks. After this correction, our utility metric is monotonic in the success rate. 
%
%
Although this does not theoretically prove a monotonic relation between utility and success rate in general, it does suggest that optimizing the utility-privacy tradeoff as in Sections \ref{sec:fundlimits} and \ref{sec:barrier} is implicitly optimizing the relation between success rate and privacy in practice.
We explore this relation further in the next section.

%% file: sections/6_numerical.tex
\section{Simulations} \label{sec:num}





In  the previous section, we have shown empirically that utility and success rate are monotonically related when deadlocks are removed.
We use this section to empirically investigate the tradeoff between success rate and privacy in simulation. 
The  purpose of these simulations is twofold: first, we want to understand more carefully  how the privacy-utility tradeoffs analyzed earlier translate into a privacy-success rate tradeoff. 
Second, we want to understand the effects of network properties  not captured by our theoretical analysis, such as graph topology,  transaction  workload, and initial capacity distribution. 

Our simulator models the sequential processing of a \emph{workload}:  a sequence of \emph{transactions}, or tuples consisting of  a sender, a receiver, and a transaction value. 
To match behavior in  real deployments, each transaction is  routed using shortest-path routing on the weighted graph, where routes are determined from public balances. 
A transaction fails if either no route can  be found, or if the  user tries a route whose true balances are insufficient to support the transaction.
When a transaction fails, we do not allow retries.
In line with theorems \ref{thm:linear} and \ref{thm:dntradeoff}, we use the all-or-nothing noise scheme $\D = \D\nve$ with optimal privacy-utility tradeoff \eqref{eqn:dntradeoff}. 
Hence, after each transaction, we update all path balances with probability $\uty(\D\nve)$, drawn independently  for each transaction.
We compute success rate by counting the fraction  of successfully-processed transactions.
Our theoretical results suggest that one can trade some privacy for an equal amount of utility. 
The question is, when we sacrifice one unit of privacy for one of utility, does this translate into a gain of more than one unit of success rate? 

Privacy-success rate curves depend on many factors, including the choice of network, workload, noise mechanism, and initial balance/capacity allocations. 
For each parameter setting, we generated \blue{50} sets of networks and/or workloads in advance and  ran each workload on  its corresponding network, with the appropriate noise mechanism.
We plot the mean success rate and standard error bars. 







We looked into the effects of network structure, transaction value, workload size and network size on performance. We focus on network structure in this section and defer the other factors to Appendix \ref{app:sims}.

%
We consider both synthetic and real network structures. 
For real network topologies, we use a snapshot of the \textbf{Lightning Network} and its channel capacities measured on December 28, 2018. 
\wt{Its major connected component} consists of 2,266 nodes and 15,392 channels, while channel capacities vary from 1,100 to 16,777,216$=2^{24}$ satoshis (1 satoshi = $10^{-8}$ Bitcoin). 
\wt{For synthetic structures, we use four different options with equal number of nodes and expected number of edges to LND:
1) \ert{} graph with expected density $6.00 \times 10^{-3}$.
2) $\ba(\ba(K_3, 113, 3), 2150, 7)$. Recall that $\ba(\texttt{init}, n, m)$ denotes a random graph initialized as \texttt{init}, to which $n$ new nodes are added sequentially. On addition of each node, $m$ new links to existing nodes are created.
3) A star-like network based on the user-server model where the server network is $\ba (K_{140}, 26, 137)$, and each user is attached to a server with probability proportional to server's degree. 
4) An LND-like network. It is initialized as $\ba ( \ba (K_8, 75, 5), 965, 13)$, where degrees of nodes are relatively high $( \ge 5)$. Then, each low-degree node is inserted and attached to 1, 2, 3 or 4 high-degree nodes uniformly at random. Eventually, it has equal number of nodes with degrees 1, 2, 3 and 4 to LND. This network is chosen to support our hypothesis about connection between success rate and number of low-degree nodes. }
\apdx{We also explore the effects of varying network size in Appendix \ref{app:sims}. }


In order to inspect the effects of topology, we set the capacity of every channel to be 1,000, and uniformly allocated balances on both ends of each channel. 

For the transaction workload, there is no canonical dataset available, so we uniformly and independently drew 2 nodes as the sender and receiver of each transaction, with transaction values independently following a Pareto distribution $\pareto(1.16, 1000)$. 
We chose a Pareto distribution to make the transaction value distribution  heavy-tailed, and selected parameters corresponding to the Pareto principle. 
$\pareto(\beta, v_m)$ has CDF
\[ F(v) = \brb{1 - \brb{\dfrac{v_m(\beta-1)}{\beta v}}^\beta}\cdot \I\brb{v \ge \frac{v_m(\beta-1)}{\beta}}. \]
Random variable $V \sim \pareto(\beta, v_m)$ has mean $\E[V] = v_m$. 
So our expected transaction value and channel capacities equal 1,000 sat. 

\begin{figure}[t]
    \centering
    \includegraphics[width=\mywidth]{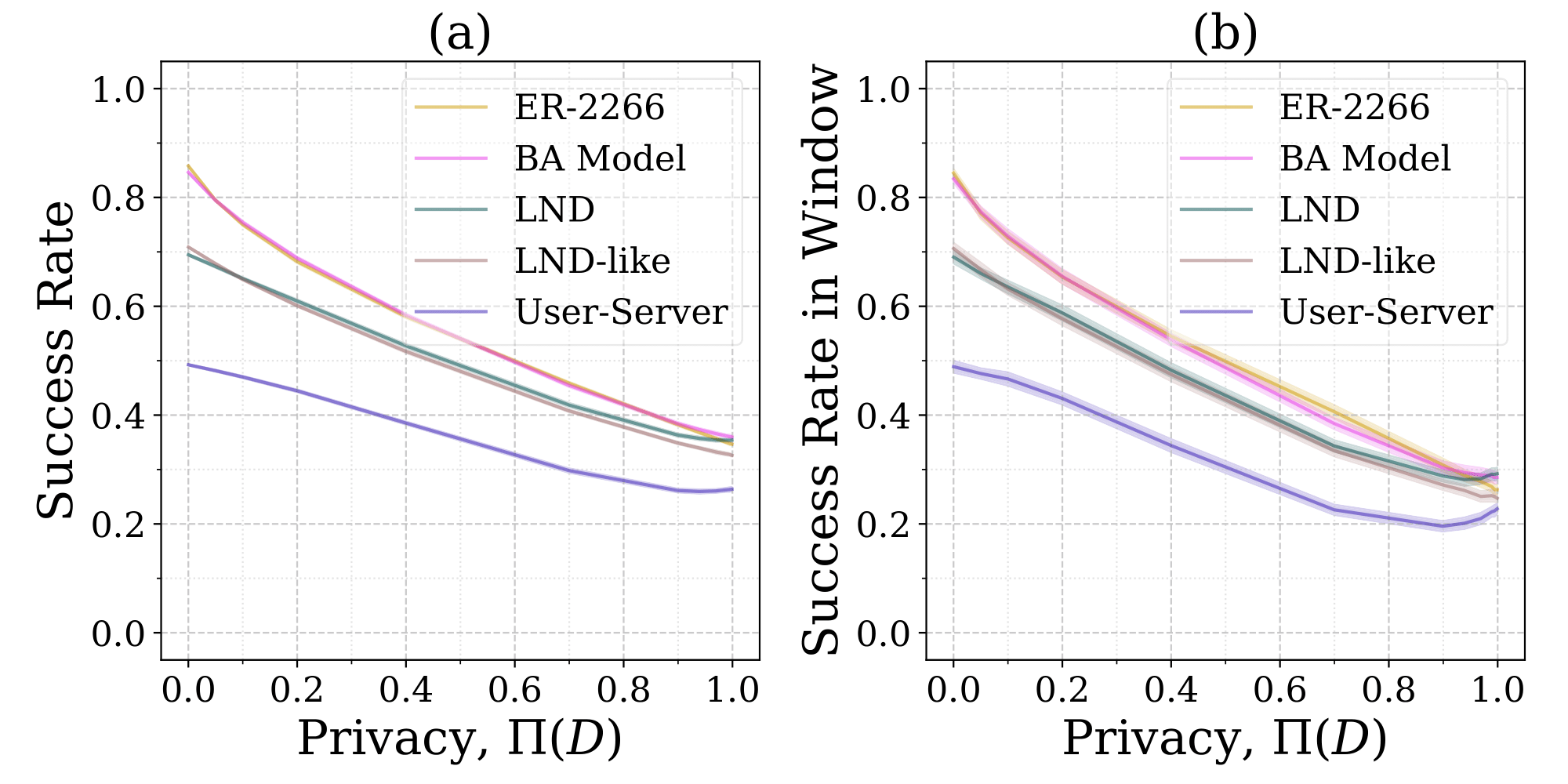}
    \caption{\blue{ Success rate-privacy curves on different network topologies. Success rate is the fraction of successful transactions out of total 100,000, while ``success rate in window'' is the fraction out of 2,000 most recent transactions. }} 
    \label{fig:exptop}
\end{figure}

The privacy-success rate tradeoff curves with error bars are shown in Figure \ref{fig:exptop}. 
The width of error bars suggests a low standard error. 
Notice three trends:

\wt{
Our first observation is the main takeaway message of this plot: \textbf{given a privacy not close to 1, we have the success rate hierarchy $\text{User-Server} < \text{LND} \approx \text{LND-like} < \text{ER} \approx \text{BA}$, 
coinciding with the reverse order of numbers of leaf nodes, which suggests a strong negative correlation between them.
}
This makes intuitive sense if we look into a specific leaf node in a PCN. 
Let privacy \revise{be} 0 (i.e., noise is not applied), and Alice be a leaf node in the PCN, who has no tokens left on her only channel. We only consider transactions involving her, i.e., which she sends, receives, or relays. 
As a leaf node, she can't relay any transaction. 
Because she has no tokens, all of her attempts to send tokens out will fail, unless another node sends tokens back to her, rebalancing her channel.
If she had a second channel, she would have more chances to successfully send tokens out. 
In addition, she would be able to relay transactions, which makes more flexible her connectivity to other nodes. 
This trend seems to remain when utility is not perfect, except when it's close to 0 and deadlocks emerge with higher probability.
Noticing the user-server model has the lowest success rate curve, it is necessary for system designers to think of an addition trade-off between a gain in privacy (as proved in Section \ref{sec:usm}) and a loss in success rate, if choice of topology is available.

%
%

Second, "success rate in window" on Figure \ref{fig:exptop}(b) has only a slight difference from success rate on Figure \ref{fig:exptop}(a).
Since the PCN is modelled as a Markov chain, both success rates will equal the probability of success of a next random transaction, when the Markov chain reaches its stable state. 
Naturally, the ``success rate in window'' taken from most recent 2,000 transactions could be an estimator of success rate at some point in the future. 
So we may predict that the non-monotonicity on Figure \ref{fig:exptop}(b) will eventually appear on the success rate-privacy figure on Figure \ref{fig:exptop}(a). 

Third, on both sub-figures of Figure \ref{fig:exptop}, if privacy is far enough away from 1, success rate decreases as privacy increases;
but on Figure \ref{fig:exptop}(b), when privacy approaches 1, success rate may increase along with privacy. 
This observation remains consistent with "original" curves in Figure \ref{fig:alleviation}. 
However, noticing the "increase" phase is much less significant than \revise{in} Figure \ref{fig:alleviation}, we may assume deadlocks have much less effect on success rate of large-scale PCNs. 
Even without alleviation, a monotonic relation between success rate and utility (or privacy, of all-or-nothing noise) can be assumed.


%
%
%
%
} 

\paragraph{Tradeoffs}
Recall that one of our main goals was to understand if the one-to-one tradeoff of privacy for utility extends to success rate. 
Figure \ref{fig:exptop} suggests that this is not the case. 
For example, on the LND curve in  Figure \ref{fig:exptop}(a), if we start at today's operating point of perfect privacy, 
 one must decrease privacy by \blue{0.4} to gain 0.1 units of success rate. 
More generally, the curves in these figures are convex with low curvature; sacrificing some amount of privacy gives disproportionately small gains in success rate. 
This suggests that system designers must sacrifice significant amounts of privacy to achieve meaningful improvements in success rate. 
\wt{In addition, it's worth paying attention to the non-monotonicity of ``success rate in window'' in Figure \ref{fig:exptop}(b). 
With a sufficiently large workload, one can always observe the decrease of success rate when privacy is sacrificed a little from perfect level. 
Hence system designers should be wary of deploying a noise mechanism that sacrifices only a little privacy; 
not only are the gains modest in theory, they can actually be negative in practice, unless system designers deal with deadlocks.}


%% file: sections/7_related.tex
\section{Related Work} \label{sec:rw}
\blue{At a high level, our problem resembles privacy-preserving routing problems that have received substantial attention over the years \cite{chaum1988dining,reiter1999anonymous,goel2003herbivore,fanti2015spy,fanti2017hiding,tsitsiklis2018delay,erturk2019dynamically,freedman2002tarzan,dingledine2004tor,capkun2004secure,nandi2011p3}; however, PCNs introduce problem constraints that preclude the use of prior algorithms and analysis.     
For instance, early systems for receiver-anonymous messaging transmitted messages to additional nodes to hide the true destination \cite{chaum1988dining,goel2003herbivore,sherwood2005p5}. 
This approach does not work in PCNs because messages are literally money, so sending a payment to additional nodes multiplies the cost of the transaction (the additional routing fees incurred by using longer paths are generally small in comparison).
In other work \cite{tsitsiklis2018delay,erturk2019dynamically}, a message is routed to prevent an adversarial observer from guessing the intended receiver prior to the message's delivery.
This problem formulation is incompatible with  PCNs, as all channel balances on the path are simultaneously updated only after a transaction has cleared.
Moreover, the main concern in  PCNs is that an adversary can infer the endpoints of a path \emph{after} execution,  not in an online setting. 
Finally, papers on source hiding in one-to-many  broadcast settings do not apply to the point-to-point routing of PCNs, which updates balances only on a single path between two users \cite{chaum1988dining,fanti2015spy,fanti2017hiding}. 
}

\blue{Another relevant line of work analyzed the robustness of communication networks to adversarially-chosen corrupted links \cite{cunningham1985optimal,gueye2010design,gueye2011network}. 
Roughly, these formulations can be interpreted as computing the worst-case success rate of a PCN following one or more adversarially-chosen transactions (though they  do not consider privacy at all). 
A key difference is that in \cite{cunningham1985optimal,gueye2010design,gueye2011network}, the adversary directly corrupts a single edge at a time, whereas in a PCN, a transaction alters \emph{all} the link balances in its path, and successive transactions can amplify imbalance. 
This prevents the analysis in \cite{gueye2010design,gueye2011network} from applying directly. 
However, this line of work suggests an interesting question: if one were to add a single channel to a PCN to maximize throughput under worst-case workloads, what channel should one add?
This question is left for future work.
}

In parallel, PCNs have been increasingly studied by the academic community, typically focusing either on privacy or on utility. 
To our knowledge, our work is the first to explicitly quantify tradeoffs between privacy and utility in PCNs. 

On the utility side, papers have primarily focused on improving the success rates of PCNs.
This is typically achieved through new routing mechanisms, which exploit ideas from the networking literature, such as packet-switched routing, congestion control, and/or flow control \cite{sivaraman2018routing,sivaraman2019high,dong2018celer}.
Our work differs in assuming source-routed transactions, both for analytical tractability and because (almost) all existing implementations of PCNs today use source routing \cite{lightning,raiden}. 
Other recent utility work has instead focused on related but orthogonal issues, such as ensuring liveness in PCNs \cite{werman2018avoiding,malavolta2017concurrency} and rebalancing depleted channels  \cite{khalil2017revive}.

On the privacy front, several papers have explored mechanisms for privacy-preserving PCN transaction routing \cite{malavolta2017silentwhispers,roos2017settling}.
There are two key differences between these papers and our work. 
First, our adversary is a passive observer of the information publicly released to all participants, 
whereas prior work like \cite{malavolta2017silentwhispers,roos2017settling,dingledine2004tor,freedman2002tarzan} considers a corrupt relay node that is trying to learn the destination of a transaction.
Malavolta \emph{et al.} \cite{malavolta2017concurrency} studied \emph{passive value privacy}, or hiding the amount of a transaction to a passive external observer; 
they also study \emph{active relationship anonymity}, which aims to hide the participants of a transaction to an active router. 
Our model is a combination of these; it considers a passive observer that wants to infer the endpoints of a transaction. 
Hence, our adversarial model is weaker than prior work in the sense that it sees less information than a router node, but stronger in the sense that encrypting the destination of a transaction does not solve the problem. 
Our results are already negative, so with a stronger active adversary, even worse tradeoffs could emerge. 
Second, our privacy problem is fundamentally different from that addressed in prior work. 
Because all network participants must be able to route transactions, encrypting transaction balances is not a viable solution; 
statistical noise is therefore more natural.
The most closely related work to ours (\cite{malavolta2017concurrency}) explicitly does not consider the problem of path selection, focusing instead on how to execute transactions once a path is selected. 



Notice that differential privacy (a common statistical privacy metric) is difficult to implement in our setting \cite{dwork2011differential}. 
For example, consider providing $\epsilon$-local differential privacy on the source of a transaction \cite{kairouz2014extremal}. 
This would require that for any observed path trace $Q$ and any pair of candidate source nodes $u,v\in \mathcal V$, it must hold that $\P[Q|s(P)=v]\leq e^\epsilon \P[Q|s(P)=u]$.
To achieve this condition, we would need to add noise to edges that are not involved in the transaction itself; 
this adds significant overhead, both in terms of implementation and in terms of added noise. 
Like differential privacy, our metric is worst-case; however, it does not require hiding the transaction participants among all nodes in the graph.

%% file: sections/8_discussion.tex
\section{Discussion}
In this paper, we have shown that for an important class of routing mechanisms (shortest-path routing), it is not possible to obtain a good privacy-utility tradeoff by releasing noisy channel balances in PCNs. 
The diagonal bound places an upper limit on the privacy-utility tradeoff obtainable in practice, and it is tight for general graph topologies. 
Our theoretical results are backed by simulations, which show that even if one considers a more complex end utility metric (i.e.~success rate), the tradeoff does not improve; in fact, it appears to worsen in some cases. 
For example, we observe a new deadlocking phenomenon, by which channels end up severely imbalanced in one direction, while the public balances are severely imbalanced in the other. 
These deadlocks are impossible to revert for a transaction distribution with a minimum transaction value, such as the Pareto distributions used in our simulations.
In practice, the situation may not be so dire; small transactions may be able to prevent the formation of total deadlocks.
Another key observation is that our privacy analysis considers a passive adversary.
An active adversary, which is stronger,  could cause these tradeoffs to degrade. 

Our findings suggest that network operators may not want to introduce noisy balance reporting mechanisms to trade off privacy for utility. 
Starting from today's operating point of perfect privacy, a network would need to give up almost all privacy to obtain substantial gains in success rate.
As long as users use shortest-path routing, network operators may be  better off choosing one extreme operating point or the other: perfect utility or perfect privacy. 

Despite these pessimistic conclusions, our analysis does not close the book on this issue. 
It may be possible to obtain a more promising tradeoff by modeling the effects of concurrent transactions and packet-switched routing, where transactions are split into smaller units and sent over different paths. 
Packet-switched routing can cause transactions to flow over a much larger fraction of network edges per transaction. 
This may have a similar effect to the long-path analysis in Section \ref{sec:clique} of confusing the adversary regarding the endpoints of the transaction.
Analyzing such routing mechanisms is an important and interesting question for future work. 


%% file: sections/appendix.tex
\section{Proofs}
\label{app:proofs}

We include proofs of the main results in this section. 

\subsection{Proof of Theorem~\ref{thm:linear}}

First, given that all routes are shortest paths, we prove the claim that there exists a unique pair of source and destination nodes of any path trace $Q \subseteq \ec$. Let $\delta(u, v)$ denote the distance between any pair of nodes $u, v$ on the graph. 
Assume by contradiction that $Q$ is a path trace from $x$ to $y$, and from $z$ to $w$, where $\brc{x, y} \neq \brc{z, w}$. By definition, there exist paths $P_{zw}$ and $P_{xy}$, where
\begin{enumerate}
    \item The endpoints of $P_{zw}$ are $z$ and $w$;
    \item The endpoints of $P_{xy}$ are $x$ and $y$;
    \item $Q$ is a subset of both $P_{zw}$ and $P_{xy}$.
\end{enumerate}

As a result, both $x$ and $y$ are on path $P_{zw}$. Because $P_{zw}$ is the shortest path connecting $z$ and $w$, its segment connecting $x$ and $y$ is also the shortest path connecting $x$ and $y$. Since $\brc{x, y}$ is different from $\brc{z, w}$, we have $\delta(x, y) < \delta(z, w)$.
By the same reasoning, we get $\delta(z, w) < \delta(x, y)$ from $z, w$ belonging to path $P_{xy}$, which is a contradiction. 
Therefore, any path trace $Q \subset \ec$ with a source has a unique source. 
So there exists a function $s:2^{\ec}-\{\emptyset\} \to \vc$, mapping each path trace to its source. Define $s(\emptyset) \notin \vc$.

To upper bound our privacy metric, we will analyze the privacy achievable by an adversary with strategy $\A\nve$, where 
\begin{enumerate}
    \item If no edge is updated by the noise mechanism, the adversary guesses randomly, i.e. $\A\nve[v|\emptyset] = \frac{1}{n}$ for all $v \in \vc$;
    \item If a set of edges $Q\neq\emptyset$ is updated, the adversary estimates its source, i.e. 
    $\A\nve[s(Q)|Q] = 1$.
\end{enumerate}

Let $\epsilon_s(P)$ denote the first edge of path $P$. By our privacy definition,
\begin{align}
    \prv(\D) &= 1 - \sup_{\A} \min_{P \in \pc} \sum_{Q \subset P} \pol{P}{Q} \adv{Q}{\partial P} \notag \\
    & \le 1 - \min_{P \in \pc} \sum_{Q \subset P} \pol{P}{Q} \A\nve[\partial P|Q]\notag \\
    & = 1 - \min_{P \in \pc} \brb{\frac{2\pol{P}{\emptyset}}{n} + \sum_{Q:s(Q) = s(P), Q \subset P} \pol{P}{Q}}, \label{eqn:diagpf}
\end{align}
where the last equality comes from splitting the adversary's probability of success into random guessing if no edges are updated, plus the probability of the noise mechanism revealing a path trace that truthfully reveals the true source edge. 
Recall that $\adv{Q}{\partial P}$ denotes the adversary's probability of guessing one of the endpoints of $P$ upon observing path trace $Q$.

Because $\D[\cdot|P]$ is a probability distribution, the following constraints hold.
\begin{align*}
    \D[\emptyset|P] &\ge 0; \\
    \sum_{Q: s(Q) = s(P), Q \subset P} \D[Q|P] = \sum_{Q: Q \ni \epsilon_s(P), Q \subset P} \D[Q|P] & \ge \uty(\D),
\end{align*}
where the latter condition follows from the definition of utility. 
If we assume $n \ge 2$, it follows that for any $P$,
\[ 
    \frac{2\pol{P}{\emptyset}}{n} + \sum_{Q:s(Q) = s(P), Q \subset P} \pol{P}{Q} 
    \ge \revise{U(\D)}. 
\]

By substituting it back to \eqref{eqn:diagpf}, we obtain
\[ 
    \Pi(\D) \le
    \revise{1 - U(\D)},
\]

\revise{which concludes the proof.} \qedb

\subsection{Proof of Proposition~\ref{thm:usm}}

For path $P \in \pc$, let $\bar P$ denote $P$ excluding the user-server channels, if they exist. If the all-or-nothing noise $\D\nve$ selects "nothing", then the adversary sees no update. Otherwise, the adversary will see $\bar P$ instead of $P$ in our original model analyzed in Section \ref{sec:fundlimits}. By definition of privacy,
\begin{align*}
    \prv(\D\nve) 
    =&~ 1 - \sup_{\A} \min_P \brc{\A[\partial P|\bar{P}] \alpha + \A[\partial P|\emptyset] \bra{1 - \alpha} } \\
    =&~ 1 - \sup_{\A[\cdot|\bar P], \A[\cdot|\emptyset]} \min_P \brc{\A[\partial P|\bar{P}] \alpha + \A[\partial P|\emptyset] \bra{1 - \alpha} } \\
    \stackrel{(*)}{\le} &~ 1 - \sup_{\A[\cdot|\bar P], \A[\cdot|\emptyset]} \brb{\alpha\min_P \A[\partial P|\bar{P}]  + \bra{1 - \alpha} \min_{P'} \A[\partial P'|\emptyset] } \\
    = &~ 1 - \alpha \sup_{\A[\cdot|\bar P]} \min_P \A[\partial P|\bar{P}]  - \bra{1 - \alpha} \sup_{\A[\cdot|\emptyset]} \min_{P'} \A[\partial P'|\emptyset], 
\end{align*}
where the last line follows because $\A[\cdot|P]$ and $\A[\cdot|\emptyset]$ are independent. 
Because the network is reachable, we may apply Lemma \ref{thm:unifguess} and obtain 
\[ \sup_{\A} \min_{P'} \A[\partial P'|\emptyset] = \frac{2}{n_S + n_U}, \]

where the adversary reaches the supremum by choosing each node with equal probability when it sees no balance update. 

Now we compute $\sup_{\A} \min_P \A[\partial P | \bar P]$. Since $\bar P$ must be a path between server nodes, we let $x, y$ denote them and define two sets 
\[ X \triangleq \brc{x} \cup \brs{v \in \vc_U}{J(v) = x},~~ Y \triangleq \brc{y} \cup \brs{v \in \vc_U}{J(v) = y}. \]

If $x = y$, then the transaction only goes through user-server channels and $\bar P = \emptyset$. The adversary will see no update regardless of utility $\alpha$, so random guessing with equal probability is the best adversarial strategy. 
Now we assume $x \neq y$, and immediately $X \cap Y = \emptyset$. The two endpoints of $P$ must lie in each of $X$ and $Y$. Because $\A[\cdot | \bar P]$ is a probability distribution, for any $\A$ there exists $u \in X$ and $v \in Y$, where 
\[ \A[u |\bar P] + \A[v | \bar P] \le \max\brc{\frac{1}{\abs{X}},~ \frac{1}{\abs{Y}}}. \]

By the properties of a user-server model, there exists a path $P_u$ containing $\bar P$, and connecting $u$ and $v$. Hence, we have 
\begin{equation}
    \sup_{\A} \min_{P} \A[\partial P' | \bar P'] \le \sup_{\A} \A[\partial P_u | \bar{P}] \le \max\brc{\frac{1}{\abs{X}},~ \frac{1}{\abs{Y}}} \le \frac{1}{\uu + 1}. \label{eqn:usmineq}
\end{equation}

Recall that $\A\sv$ was previously defined as $\A\sv[v|Q] = 1/(n_S + n_U)$ for all $v \in \vc$, if $Q = \emptyset$. 
We supplement this definition by defining it also in the case $Q \neq \emptyset$. 
Recall that all-or-nothing noise $\D\nve$ is being used, and $Q$ must be a shortest path on the server network $(\vc_S, \ec_S)$ from $x'$ to $y'$. Similar to $X$ and $Y$, we use the following notations.
\[ X'= \brc{x'} \cup \brs{v \in \vc_U}{J(v) = x'},~ Y'= \brc{y'} \cup \brs{v \in \vc_U}{J(v) = y'}. \]

Assume $Z'$ is the set with fewer elements between $X'$ and $Y'$. The strategy $\A\sv$ is defined as followed.
\[ \A\sv[z | Q] = \left\{ \begin{array}{lr} \frac{1}{\abs{Z'}}, & \text{if } z \in Z',\\ 0, & \text{otherwise.} \end{array} \right. \]

It can be easily confirmed that $\A\sv$ can make every inequality in \eqref{eqn:usmineq} tight, thus \[ \sup_{\A} \min_P \A[\partial P | \bar P] = \frac{1}{\mu+1}. \]

By substituting $\A = \A\sv$ in both sides of (*), we also get an equality. 
Hence,
\[ \prv(\D\nve) = 1 - \frac{1}{\mu+1} \alpha - \bra{1 - \alpha} \frac{2}{n_S + n_U}, \]

which proves \eqref{eqn:usm}. \qedb

\subsection{Proof of Proposition \ref{thm:usmv2}}

For path $P \in \pc$, let $\bar P$ denote $P$ excluding the user-server channels, if they exist. Furthermore, let $\qc = \brs{\bar P}{P \in \pc}$. By definition of privacy,
\begin{align*}
    & ~ \prv(\D\nve) \\
    =&~ 1 - \sup_{\A} \min_P \brc{\A[\partial P|\bar{P}] \alpha + \A[\partial P|\emptyset] \bra{1 - \alpha} } \\
    =&~ 1 - \sup_{\A[\cdot|\neg \emptyset], \A[\cdot|\emptyset]} \min_P \brc{\A[\partial P|\bar{P}] \alpha + \A[\partial P|\emptyset] \bra{1 - \alpha} } \\
    \stackrel{(*)}{\le} &~ 1 - \sup_{\A[\cdot|\neg \emptyset], \A[\cdot|\emptyset]} \brb{\alpha\min_P \A[\partial P|\bar{P}]  + \bra{1 - \alpha} \min_{P'} \A[\partial P'|\emptyset] } \\
    = &~ 1 - \alpha \sup_{\A[\cdot|\neg \emptyset]} \min_P \A[\partial P|\bar{P}]  - \bra{1 - \alpha} \sup_{\A} \min_{P'} \A[\partial P'|\emptyset] \\
    = &~ 1 - \alpha \sup_{\A[\cdot|\neg \emptyset]} \min_P \A[\partial P|\bar{P}] - \frac{2}{n}(1-\alpha) \\
    = &~ 1 - \frac{2}{n}(1-\alpha) - \alpha \min_{Q \in \qc} \sup_{\A[\cdot|Q]} \min_{P: \bar P = Q} \A[\partial P|Q]. \tag{**} \label{eqn:usmv2prv}
\end{align*}

Note that by Lemma \ref{thm:unifguess}, $\sup_{\A} \min_{P'} \A[\partial P'|\emptyset] = 2/n$. Any adversarial strategy can achieve this supremum by guessing uniformly at random when seeing no updates. 

For a server node $v$, we define its \textbf{cloud} $C_v$ as below. 
$$C_v = \{v\} \cup \brs{u \in \vc_U}{(u, v) \in \ec_U}. $$ 

Apparently, for all $v \in \ec_S$, $\abs{C_v} \ge \uu + 1$. Let $x, y \in \vc_S$ where $x \neq y$. Since $C_x$ and $C_y$ are not disjoint anymore, we define 
\[ Z = C_x \cap C_y, ~~ X = C_x - Z,~~ Y = C_y - Z. 
\]

\newcommand{\ax}{\abs{X}}
\newcommand{\ay}{\abs{Y}}
\newcommand{\az}{\abs{Z}}

Apparently, $X, Y$ and $Z$ are disjoint subsets with union $C_x \cup C_y$. In addition, $x \in X$, $y \in Y$. For now, we assume $\az \ge 2$, so that there exists a path whose both source and destination belong to $Z$. 

Because the network is reachable, there exists a path $Q$ on the server network $(\vc_S, \ec_S)$ connecting $x$ and $y$. In addition, for any $x' \in C_x$ and $y' \in C_y$ where $x' \neq y'$, there exists a path $P$ with endpoints $x', y'$, and containing $Q$. In this case, we have
\[ \min_{P} \A\sv[\partial P|Q] = \min\brc{p_X+p_Y, p_X+p_Z, p_Y+p_Z, 2p_Z}, \]
where for $W \in \brc{X, Y, Z}$, $p_W \triangleq \min_{v \in W} \A\sv[v|Q]$. Now obviously, the best adversary $\A\sv$ satisfies for all $x' \in X,~ y'\in Y,~ z'\in Z$,
\[ (\A\sv[x'|Q],~ \A\sv[y'|Q],~ \A\sv[z'|Q]) \equiv \bra{p_X,~ p_Y, p_Z}. \]

Hence, to obtain $\A\sv[\cdot|Q]$, we only need to determine the probabilities $p_X, p_Y$ and $p_Z$ by solving the following problem.
\begin{equation}
    \label{eqn:pxyzmax}
    \begin{aligned}
        \mathrm{maximize} \quad & \min\brc{p_X+p_Y,~ p_Y+p_Z,~ p_Z+p_X,~ 2p_Z} \\
        \mathrm{subject~to} \quad & \ax p_X + \ay p_Y + \az p_Z = 1.
    \end{aligned}
\end{equation}

In order to deal with the minimum in the objective function, we discuss 4 cases.

\paragraph{Case 1: $p_Z \le p_X, p_Y$} The objective function equals $2p_Z$, and
\[ 2p_Z \le \frac{2\bra{\ax p_X + \ay p_Y + \az p_Z}}{\ax + \ay + \az} = \frac{2}{\ax + \ay + \az}. \]

By taking $p_X = p_Y = p_Z = 2/(\ax+\ay+\az)$, the upper bound is attained. So in this case the maximum equals $2/(\ax+\ay+\az)$.

\paragraph{Case 2: $p_Z \ge p_X, p_Y$} The objective function equals $p_X + p_Y$. 
\begin{itemize}
    \item If $\abs{\ax - \ay} \le \az$, we let $t = \bra{\ay - \ax + \az}/\bra{2\az}$, which lies in $[0, 1]$. Then, 
    \begin{align*}
        1 &= \ax p_X + \ay p_Y + \az p_Z \\
        &\ge \bra{\ax + t\az} p_X + \brb{\ay + (1-t)\az} p_Y \\
        &= \frac{\ax + \ay + \az}{2} \bra{p_X + p_Y}.
    \end{align*}
    
    So the maximum equals $2/(\ax+\ay+\az)$, which is attained by taking $p_X = p_Y = p_Z = 2/(\ax+\ay+\az)$.
    
    \item If $\ay > \ax + \az$, we have
    \begin{align*}
        1 &= \ax p_X + \ay p_Y + \az p_Z \\
        &\ge \bra{\ax + \az} p_X + \bra{\ax + \az} p_Y.
    \end{align*}
    
    The maximum equals $1/(\ax+\az)$, which is attained by taking $p_Y = 0$ and $p_X = p_Z = 1/(\ax+\az)$.
    
    \item If $\ax > \ay + \az$, analogously, the maximum equals $1/(\ay+\az)$ with maximizers $p_X = 0$ and $p_Y = p_Z = 1/(\ay+\az)$.
\end{itemize}

To summarize, the maximum equals 
\[ \max\brc{\frac{2}{\ax+\ay+\az},~ \frac{1}{\ax+\az},~ \frac{1}{\ay+\az}}. \]

\paragraph{Case 3: $p_X \le p_Z \le p_Y$} The objective function equals $p_X + p_Z$. 
\begin{itemize}
    \item If $\ax > \ay + \az$, 
    \begin{align*}
        1 &= \ax p_X + \ay p_Y + \az p_Z \\
        &\ge \bra{\ay + \az} p_X + \ay p_Z + \az p_Z.
    \end{align*}
    
    This indicates maximum of $p_X + p_Z$ equals $1/(\ay+\az)$ by picking $p_X = 0$ and $p_Y = p_Z = 1/(\ay+\az)$.
    
    \item If $\ax \le \ay + \az$, 
    \begin{align*}
        1 &= \ax p_X + \ay p_Y + \az p_Z \\
        &\ge \ax p_X + (\ay + \az) p_Z \\
        &\ge \frac{\ax+\ay+\az}{2}\bra{p_X + p_Z}.
    \end{align*}
    
    The last line is obtained by using Chebyshev's sum inequality, with conditions $\ax \le \ay + \az$ and $p_X \le p_Z$. So the maximum of $p_X + p_Z$ equals $2/\bra{\ax+\ay+\az}$, which is attained at $p_X = p_Y = p_Z = 1/\bra{\ax+\ay+\az}$.
\end{itemize}

To summarize, the maximum equals 
\[ \max\brc{\frac{2}{\ax+\ay+\az},~ \frac{1}{\ay+\az}}. \]

\paragraph{Case 4: $p_Y \le p_Z \le p_X$} Analogously, the maximum equals 
\[ \max\brc{\frac{2}{\ax+\ay+\az},~ \frac{1}{\ax+\az}}. \]

In general, the maximum may be simply expressed as below.
\begin{equation}
    \label{eqn:pxyzsol}
    \sup_{\A[\cdot|Q]} \min_{P} \A[\partial P|Q] = \max\brc{\frac{2}{\ax+\ay+\az}, \frac{1}{\ax+\az}, \frac{1}{\ay+\az}}.
\end{equation}

To attain it, the optimal adversarial strategy $\A\sv$ satisfies 
\begin{enumerate}
    \item if $\abs{\ax-\ay} \le \az$, then for all $v \in X\cup Y\cup Z$, $\A\sv[v|Q] = 1/\bra{\ax+\ay+\az}$;
    \item if $\ax > \ay + \az$, then for all $x' \in X$ and $v \in Y \cup Z$, $\A\sv[x'|Q] = 0,~ \A\sv[v|Q] = 1/\bra{\ay+\az}$;
    \item if $\ay > \ax + \az$, then for all $y' \in Y$ and $v \in X\cup Z$, $\A\sv[y'|Q] = 0,~ \A\sv[v|Q] = 1/\bra{\ax+\az}$.
\end{enumerate}

It should be addressed that the above solution applies only when $\az \ge 2$. For the case $\az = 0$, we have $C_x \cap C_y = \emptyset$ equivalently. Proof of Proposition \ref{thm:usm} has shown that the maximum value and the optimal strategy are actually compatible with \eqref{eqn:pxyzsol}. Hence, we only need to focus on the case $\az = 1$. The analysis begins to fork at the problem formulation, because there will be no path with both ends lying in set $Z$. So we need to modify \eqref{eqn:pxyzmax} as followed. 
\begin{equation}
    \label{eqn:pxyzmaxz1}
    \begin{aligned}
        \mathrm{maximize} \quad & \min\brc{p_X+p_Y,~ p_Y+p_Z,~ p_Z+p_X} \\
        \mathrm{subject~to} \quad & \ax p_X + \ay p_Y + \az p_Z = 1.
    \end{aligned}
\end{equation}

After similar discussions, we can obtain the maximum
\[ \max\brc{\frac{2}{\ax+\ay+\az}, \frac{1}{\ax+\az}, \frac{1}{\ay+\az}, \frac{1}{\ax + \ay}}. \]

Because $\az = 1 \le \ax, \ay$, both \eqref{eqn:pxyzsol} and $\A\sv$ apply to this case. So we may regard them as a general solution, without respect to the value of $\az$. Recalling $\uu$ is the minimum number of user nodes connected to any server,
\begin{align*}
    & \max\brc{\frac{2}{\ax+\ay+\az}, \frac{1}{\ax+\az}, \frac{1}{\ay+\az}} \\
    =~ & \max\brc{\frac{2}{\abs{C_x \cup C_y}}, \frac{1}{\abs{C_x}}, \frac{1}{\abs{C_y}}} \\
    \le ~ & \max\brc{\frac{2}{\uu+2}, \frac{1}{\uu+1}, \frac{1}{\uu+1}} 
    = \frac{2}{\uu+2}.
\end{align*}

Now we go back to privacy. It's easy to confirm that by substituting the supremum over $\A$ with $\A\sv$ at both sides of (*), it turns into an equality. Following \eqref{eqn:usmv2prv},
\begin{align*}
      \prv(\D\nve) 
    =~ & 1 - \frac{2}{n}(1-\alpha) - \alpha \min_{Q \in \qc} \sup_{\A[\cdot|Q]} \min_{P: \bar P = Q} \A[\partial P|Q] \\
    \ge~ & 1 - \frac{2}{n}(1-\alpha) - \alpha \min_{Q \in \qc} \frac{2}{\uu+2} \\
    =~ & \bra{1-\frac{2}{n}}\bra{1-\alpha} + \frac{\uu}{\uu+2} \alpha.
\end{align*}

So the lower bound of privacy is justified. \qedb

\subsection{Proof of Theorem  \ref{thm:alttradeoff}} 
Let 
$\ell = \floor{L/2}$. 
Let $\bar P_1$ and $\bar P_2$ denote the "odd" group and "even" group of path $P$, respectively. 
We define 
\begin{align*}
    \qc_i &\triangleq \brs{\bar P_i}{P \in \pc},~ i \in \brc{1, 2},\\
    Z(v, Q) &\triangleq \abs{\brs{P}{v \in \partial P,~ Q \in \brc{\bar P_1, \bar P_2}}}.
\end{align*}

The method to analyse $Z(v, Q)$ 
depends on the parity of $L$.

 \paragraph{Case 1:}  When $2 \mid L$, we have $\qc_1 = \qc_2$ (because the graph is complete). 
    For convenience and consistency, we introduce some terminology  that will be reused in the proof of Theorem \ref{thm:iidc}.
    For any path trace $Q$  of $P$, it consists of several disconnected segments, where each segment consists of one or more edges connected end to end (in  the alternating noise scheme, each segment is only one edge). 
    Each segment has two end nodes and zero or more intermediate nodes. 
    We call the end nodes of a segment \textbf{gray nodes}, and the intermediate nodes are called \textbf{white nodes}. 
    Again, in the alternating noise scheme, there are no white nodes; we will use white nodes in the proof of Theorem \ref{thm:iidc}.
    The rest of the nodes in the network are called \textbf{black nodes}. 
    Expressing these as sets, we let $g_Q, w_Q$ and $b_Q$ denote the sets of gray, white and black nodes given path trace $Q$. 
    An example of segments and the aforementioned colors are shown in Figure \ref{fig:clqsegment}. 
    
    For any path $P$ under the alternating noise scheme, both $\bar P_1$ and $\bar P_2$ yield $L=2\ell$ gray nodes and 1 black node.
    All nodes not on path $P$ are always black nodes.
    If a gray node $v$ is determined to be one endpoint of $P$, then its status as a source or destination is determined too. 
    In that case, the other end must be a black node. 
    So $Z(v, Q)$ equals the number of ways to pick this black node, multiplied by the number of ways to permute the remaining $\ell-1$ segments. 
    If a black node $v$ is determined to be one end, it could be either a source or a destination. Then, $Z(v, Q)$ equals the number of ways to permute all $\ell$ segments, multiplied by $2$, the number of ways to determine which end $v$ is. That is,
    \[ Z(v, Q) = \left\{ \begin{array}{lr}
        (\ell-1) ! (n-2\ell), & v \in g_Q; \\
        2\ell !, & v \in b_Q.
    \end{array} \right. \]
    
\paragraph{Case 2:} When $2 \nmid L$, the case is different because for all $Q\in \qc_1$, $Q$ has $\ell+1$ edges, while for all $Q \in \qc_2$, $Q$ has $\ell$ edges. As a result, $\qc_1 \cap \qc_2 = \emptyset$. For all $Q \in \qc_1$,
    \[ Z(v, Q) = \left\{ \begin{array}{lr}
        \ell !, & v \in g_Q; \\
        0, & v \in b_Q.
    \end{array} \right. \]
    
    For all $Q \in \qc_2$, 
    \[ Z(v, Q) = \left\{ \begin{array}{lr}
        0, & v \in g_Q; \\
        2 \ell!(n-2\ell-1), & v \in b_Q.
    \end{array} \right. \]

\vspace{0.3in}
Define adversarial strategy $\A\alt$ as below. Note that $\A\alt[v|\emptyset] = 1/n$ for all $v \in \vc$. If balance updates of the entire path $P$ are visible, $\A\alt$ directly picks an end node of it, so for all $P \in \pc$, $\A\alt[\partial P|P] = 1$.
\begin{enumerate}
    \item When $2 \mid L$, 
    \[ \A\alt[v|Q] = \left\{ \begin{array}{lr}
    \dfrac{\I[2L<n]}{L} + \dfrac{\I[2L=n]}{n}, & v \in g_Q; \\[1ex]
    \dfrac{\I[2L=n]}{n} + \dfrac{\I[2L>n]}{n-L}, & v \in b_Q.
    \end{array} \right. \]
    
    \item When $2 \nmid L$,
    \[ \A\alt[v|Q] = \left\{ \begin{array}{lr}
    \dfrac{\I[Q \in \qc_1]}{L+1}, & v \in g_Q; \\[2ex]
    \dfrac{\I[Q \in \qc_2]}{n-L+1}, & v \in b_Q.
    \end{array} \right. \]
\end{enumerate}

Based on calculations of $Z(v, Q)$, it's easy to confirm that
\[ \A\alt[\cdot|Q] \in \argsup_{\A[\cdot|Q]} \sum_{v \in \vc} \A[v|Q] Z(v, Q),\quad \forall Q \in \qc_1 \cup \qc_2. \]

Now we analyse privacy based on the following discussions.
\begin{enumerate}
    \item When $\alpha \le 1/2$, 
    \begin{align}
        & 1 - \prv(\D\alt) \notag \\
        =~& \sup_{\A} \inf_{P \in \pc} \brc{\alpha \A[\partial P | \bar P_1] + \alpha \A[\partial P| \bar P_2] + (1-2\alpha) \A[\partial P| \emptyset] } \notag \\
        =~& \alpha \sup_{\A} \inf_{P \in \pc} \brc{\A[\partial P | \bar P_1] + \A[\partial P| \bar P_2]} + \frac{2-4\alpha}{n} \label{eqn:unifapp} \\
        \stackrel{(\dagger)}{\le}~& \frac{\alpha}{\abs{\pc}} \sup_{\A} \sum_{P \in \pc} \brc{\A[\partial P | \bar P_1] + \A[\partial P| \bar P_2]} + \frac{2-4\alpha}{n} \notag \\
        =~& \frac{\alpha}{\abs{\pc}} \sup_{\A} \sum_{Q \in \qc_1 \cup \qc_2}  \sum_{v \in \vc} \A[v|Q] Z(v, Q) + \frac{2-4\alpha}{n}. \notag
    \end{align}
    
    Note that \eqref{eqn:unifapp} is obtained by Lemma \ref{thm:unifguess}, where $\A\alt[\cdot|\emptyset]$ achieves the supremum. If we plug $\A = \A\alt$ into both sides of ($\dagger$) by replacing the supremum, we get an equality. Hence, the left side supremum of ($\dagger$) is also achieved by $\A\alt$. Therefore, we can calculate $\prv(\D\alt)$ as followed.
    
    \begin{enumerate}
        \item When $2\mid L$, let $P \in \pc$ be arbitrary. In this case, both $\bar P_1$ and $\bar P_2$ have a gray end node and a black end node. Hence,
        \begin{align*}
            & \prv(\D\alt) \\
            =~& 1 - \alpha \bra{\A\alt[\partial P|\bar P_1] + \A\alt[\partial P|\bar P_2]} - \frac{2-4\alpha}{n} \\
            =~& 1 + \frac{4\alpha -2}{n} - 2\alpha \bra{\frac{\I[2L<n]}{L} + 2\frac{\I[2L=n]}{n} + \frac{\I[2L>n]}{n-L}} \\
            =~& 1 - \frac{2}{n} - \bra{\frac{2}{\min\brc{L, n-L}}-\frac{4}{n}}\alpha.
        \end{align*}
        
        \item When $2\nmid L$, let $P \in \pc$ be arbitrary. Both end nodes are gray for $\bar P_1$ and black for $\bar P_2$. Hence,
        \begin{align*}
            \prv(\D\alt) =~& 1 - \alpha \bra{\A\alt[\partial P|\bar P_1] + \A\alt[\partial P|\bar P_2]} - \frac{2-4\alpha}{n} \\
            =~& 1 + \frac{4\alpha -2}{n} - \alpha \bra{\frac{2\I[\bar P_1 \in \qc_1]}{L+1} + \frac{2\I[\bar P_2 \in \qc_2]}{n-L+1}} \\
            =~& 1 - \frac{2}{n} - \bra{\frac{2}{L+1} + \frac{2}{n-L+1}-\frac{4}{n}}\alpha.
        \end{align*}
    \end{enumerate}
    
    \item When $\alpha > 1/2$, with probability $(2\alpha-1)$, public balances of $P$ are updated. Hence, 
    \begin{align*}
        & \prv(\D\alt)\\
        =~& 1 - \sup_{\A} \min_{P \in \pc} 
            \begin{aligned}[t]
                \big\{ & (1-\alpha)\A\alt[\partial P|\bar P_1] + (1-\alpha)\A\alt[\partial P|\bar P_2] +\\
                & (2\alpha-1) \A\alt[\partial P|P] \big\}
            \end{aligned} \\
        =~& 1 - (2\alpha-1) - (1-\alpha) \sup_{\A} \min_{P \in \pc} \brc{\A\alt[\partial P|\bar P_1] + \A\alt[\partial P|\bar P_2]}.
    \end{align*}
    
    Note that the supremum has been solved in the case $\alpha \le 1/2$. So the results will be written below directly.
    \begin{enumerate}
        \item When $2 \mid L$, 
        \[ \prv(\D\alt) = (1-\alpha) \bra{2 - \frac{2}{\min\brc{L, n-L}}}. \]
        \item When $2 \nmid L$,
        \[ \prv(\D\alt) = (1-\alpha) \bra{2 - \frac{2}{L+1} - \frac{2}{n-L+1}}. \]
    \end{enumerate}
\end{enumerate}

The tradeoffs have been proved by the discussions above. \qedb

\subsection{Proof of Theorem  \ref{thm:iidc}}

Let $\lambda=L+1$ denote the number of nodes in the path.
Define $\qc \triangleq \bigcup_{P \in \pc} 2^P$, where $2^S$ denotes the class of subset of an arbitrary set $S$. Notice that the adversarial strategy $\A$ has only one constraint -- for every $Q \in \qc$, $ \A[\cdot |Q]$ is a probability distribution over $\vc$, while $\A[\cdot|Q_1]$ is independent with $\A[\cdot|Q_2]$ for $Q_1 \neq Q_2$. Hence, we can divide $\A$ into many separate distributions with a small sample space $\vc$. 

Beginning with definition of privacy, we use a trick of comparing minimum and mean.
\begin{align}
    \prv(\D\iid) =~ & 1 - \sup_{\A} \min_{P \in \pc} \sum_{Q\subset P} \A[\partial P|Q] \D\iid[Q|P] \notag \\
    \ge~ & 1 - \sup_{\A} \frac{1}{\abs{\pc}} \sum_{P \in \pc} \sum_{Q \subset P} \A[\partial P |Q] \D\iid[Q|P] \label{eqn:minlemean} \\
    = ~& 1 - \frac{1}{\abs{\pc}} \sup_{\A} \sum_{Q \in \qc} \alpha^{\abs{Q}} (1-\alpha)^{L-\abs{Q}} \sum_{P \in \pc, P \supset Q} \A[\partial P|Q] \notag \\
    \triangleq ~& 1 - \sum_{Q \in \qc}  
    \begin{aligned}[t] & \frac{(n-\lambda)!\alpha^{\abs{Q}} (1-\alpha)^{L-\abs{Q}}}{n!}\cdot \\
    & \sup_{\A[\cdot|Q]} \sum_{v \in \vc} \A[v|Q] Z(v, Q). \end{aligned} \label{eqn:zvq} 
\end{align}

\begin{figure}
    \centering
    \includegraphics[width=.4\mywidth]{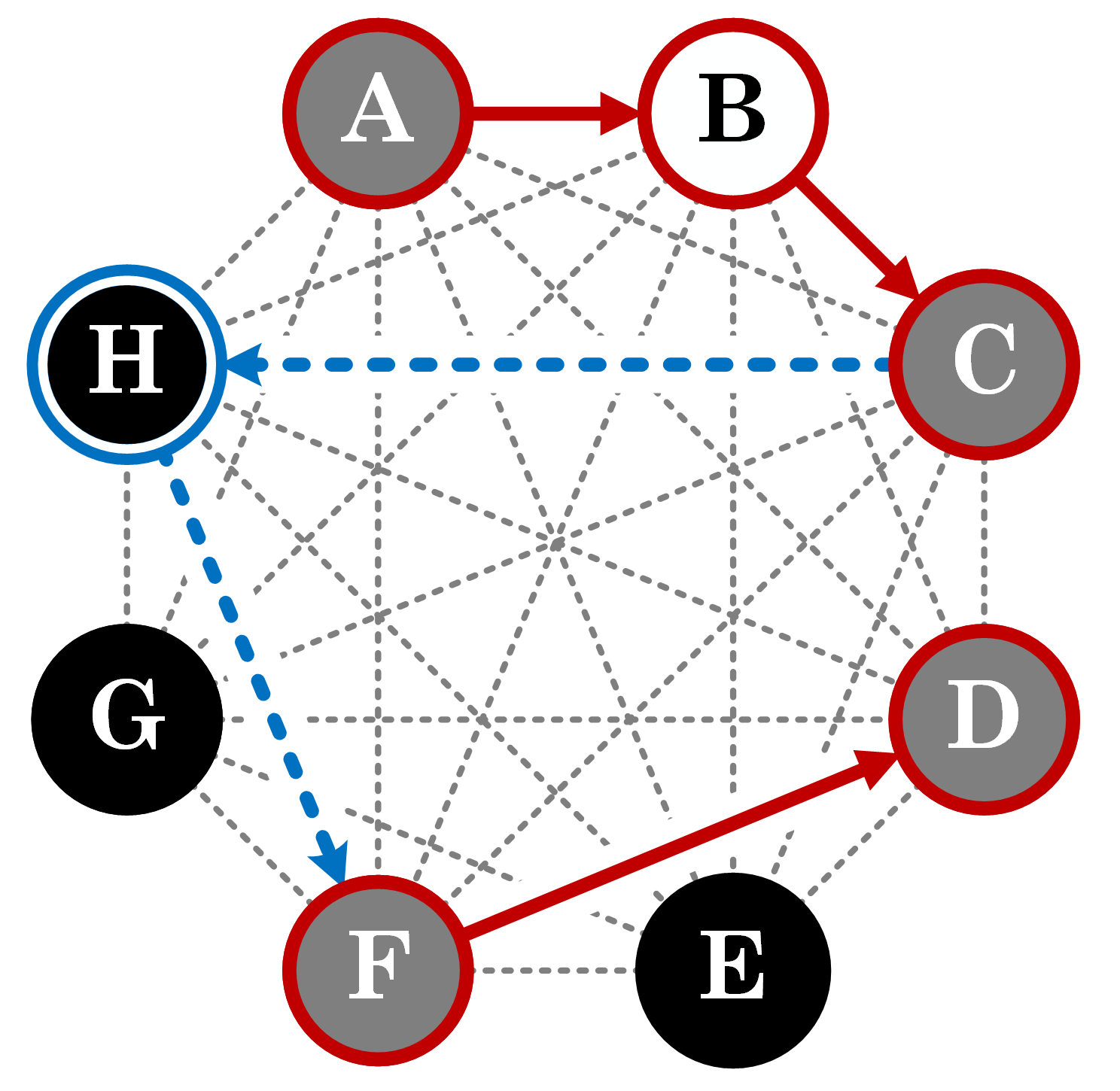}
    \caption{A path and its path trace for i.i.d. noise. Gray nodes are the endpoints of a revealed segment; white nodes are internal to a revealed segment; black nodes are not adjacent to any revealed edge. }
    \label{fig:clqsegment}
\end{figure}

\eqref{eqn:zvq} ended with substitution $Z(v, Q) = \abs{\brs{P\in \pc}{Q \subset P, v \in \partial P}}$. Since $Q$ is a path trace of $P$, it consists of several disconnected segments, where each segment consists of one or more edges connected end to end. Intuitively, a segment has two end nodes and zero or more intermediate nodes. The end nodes of segments are called \textbf{gray nodes}, and the intermediate nodes are called \textbf{white nodes}. The rest of the nodes in the network are called \textbf{black nodes}. 
Expressing these as sets, we let $g_Q, w_Q$ and $b_Q$ denote the sets of gray, white and black nodes given path trace $Q$. An example of segments and the aforementioned colors are shown in Figure \ref{fig:clqsegment}. 

The colors of nodes are decided by $Q$ consisting of disconnected segments. Assume $Q$ has $h_Q \ge 1$ segments. Furthermore, let $k_Q$ denote the number of \textit{internal} black nodes, i.e. the black nodes \textit{on the original path}. Relatively, \textit{external} black nodes are the nodes out of the original path. Now the number of white nodes equals $\lambda - 2h_Q - k_Q$ and that of external black nodes equals $n-\lambda$. The value of $Z(v, Q)$ is discussed based on the color of $v$. 

\begin{enumerate}
    \item If $v$ is a white node, it could never be an end node of a path, so $Z(v, Q) = 0$ for all $v \in w_Q$.
    \item If $v$ is a gray node, we first deal with the case that $v$ is a source node of some segment $\sigma_0$, where the other segments are denoted by $\sigma_1$ through $\sigma_{h_Q-1}$. Each original path with source node $v$ has a permutation of the rest $L$ nodes, which is uniquely mapped to a combination of the following options.
    \begin{itemize}
        \item Order of segments on the path. There are $(h_Q-1)!$ ways to permute $q_1$ through $q_{n-1}$. 
        \item Distribution of black nodes in the gaps between consecutive segments, or in the extension of the last segment. There are $\combus{h_Q+k_Q-1}{k_Q}$ different ways to distribute the slots for black nodes.
        \item Permutation of black nodes filling into the slots. Without regard of the position of the slots, there are $(n-\lambda+k_Q)!/(n-\lambda)!$ different permutations.
    \end{itemize}
    These options are independent of each other, so the multiplication rule applies. The analysis for $v$ being a destination node of a segment is identical. It follows that for all $v \in g_Q$, 
    \begin{equation}
        Z(v, Q) = \frac{(h_Q + k_Q - 1)! (n-\lambda+k_Q)!}{k_Q! (n-\lambda)!}. \label{eqn:zvqg}
    \end{equation}
    
    \item If $v$ is a black node, we must assume $k_Q \ge 1$. We first deal with the case that $v$ is a source node of the path. Similarly we discuss the following options.
    \begin{itemize}
        \item Order of segments. There are $h_Q!$ different orders.
        \item Distribution of the remaining $k_Q - 1$ slots for black nodes. There are $\combus{h_Q+k_Q-1}{k_Q-1}$ different distributions.
        \item Permutation of black nodes filling into the slots. There are $(n-\lambda+k_Q-1)!/(n-\lambda)!$ different permutations.
    \end{itemize}
    The analysis for $v$ being a destination node is identical. Considering $v$ could be both source nodes and a destination nodes of different paths, for all $v \in b_Q$,
    \begin{equation}
        Z(v, Q) = \left\{ \begin{array}{lr}
            \dfrac{2(h_Q + k_Q - 1)! (n-\lambda+k_Q-1)!}{(k_Q-1)! (n-\lambda)!}, & k_Q \ge 1; \\
            0,  & k_Q = 0.
        \end{array}\right. \label{eqn:zvqb}
    \end{equation}
\end{enumerate}

$Z(v, Q)$ depends only on the color of $v$, i.e. the set $v$ belongs to. We may use $Z_g(Q)$ and $Z_b(Q)$ to denote $Z(v,Q)$ for an arbitrary $v \in g_Q$ and $v \in b_Q$, respectively. Starting with the LP in \eqref{eqn:zvq},
\begin{align}
    &\sup_{\A[\cdot|Q]} \sum_{v \in \vc} \A[v|Q] Z(v, Q) \notag \\
    = ~& \sup_{\A[\cdot|Q]} \brb{ Z_g(Q) \sum_{u \in g_Q} \A[u|Q] + Z_b(Q) \sum_{v \in b_Q} \A[v|Q] } \notag \\
    = ~& \max\brc{Z_g(Q), ~Z_b(Q)} \notag \\
    = ~& \frac{(h_Q+k_Q-1)! (n-\lambda+k_Q)!}{(n-\lambda)! k_Q!} \psi(k_Q), \notag
\end{align}

where
\[ \psi(k_Q) = \left\{ \begin{array}{lr} 1, & k_Q \le n-\lambda; \\ \frac{2k_Q}{n-\lambda+k_Q}, & \text{otherwise}.
\end{array} \right. \]

One of the maximizers $\A\iid$ of the LP could be written as
\[ \A\iid[v|Q] = \left\{ \begin{array}{lr}
    \dfrac{\I[v \in g_Q]}{2 h_Q}, & k_Q \le n - \lambda; \\
    \dfrac{\I[v \in b_Q]}{n-\lambda+k_Q}, & \text{otherwise}.
\end{array}\right. \]

Explicitly, when the adversary sees balance updates on path trace $Q$, it picks a color from gray and black, and randomly chooses a node of that color. The choice of color depends on whether $k_Q \le n - \lambda$. Recall that $k_Q$ is the number of internal black nodes, and $n-\lambda$ is the number of external black nodes. 

Let $f(\A, P) \triangleq \sum_{Q\subset P} \A[\partial P|Q] \D\iid[Q|P]$. Based on the symmetry of $\A\iid$, it's easy to confirm that when substituting the supremum over $\A$ with $\A\iid$ in \eqref{eqn:minlemean}, it will make the inequality tight. Let $\U$ denote the uniform distribution over $\pc$. Now we obtain 
\begin{align*}
    \sup_{\A} \min_{P \in \pc} f(\A, P) & \le \sup_{\A} \E_{P\sim \U} f(\A, P) \\
    & = \E_{P\sim \U} f(\A\iid, P) = \min_{P \in \pc} f(\A\iid, P) \\
    & \le \sup_{\A} \min_{P \in \pc} f(\A, P).
\end{align*}

Hence, \eqref{eqn:minlemean} is actually an equality, and
\[ \A\iid \in \argsup_{\A} \min_{P \in \pc} \sum_{Q\subset P} \A[\partial P|Q] \D\iid[Q|P]. \]

This means $\A\iid$ is an optimal adversarial strategy for i.i.d. noise $\D\iid$ and current routing scheme. The proof will be wrapped up with a final computation of $\prv(\D\iid)$ based on the optimal $\A\iid$. Since $\min_{P \in \pc} f(\A\iid, P) = \E_{P\sim \U} f(\A\iid, P)$, for an arbitrary $P' \in \pc$, 
\[ \min_{P \in \pc} f(\A\iid, P) = f(\A\iid, P'). \] 

Immediately,
\begin{equation}
    \prv(\D\iid) = 1 - \sum_{Q \subset P'} \alpha^{\abs{Q}} (1-\alpha)^{L-\abs{Q}} \A\iid [\partial P'|Q]. \label{eqn:prvpartp}
\end{equation}

Let $v_s$ and $v_d$ denote the source and destination of path $P'$, respectively. For a particular $Q$ with $h\ge 1$ segments and $k$ internal black nodes, we have $\abs{Q} = \lambda - h - k$. Now we discuss the colors of $v_s$ and $v_d$, depending on $Q$.

\begin{enumerate}
    \item If both are black, then 
    \[ \A\iid[v_s|Q] = \A\iid[v_d|Q] = \frac{\I[k > n-\lambda]}{n-\lambda+k}. \]
    Because the path is given, the number of path traces with both black ends equals the number of arrangements of slots for black nodes and white nodes. In this case, there are $h+1$ gaps for $k-2$ black nodes, and $h$ gaps for $\lambda - 2h - k$ white nodes. Hence,
    \[ \abs{\brs{Q \subset P}{v_s, v_d \in b_Q}} = \combus{h+k-2}{k-2} \combus{\lambda-h-k-1}{h-1}. \]
    
    \item If $v_s$ is black and $v_d$ is gray, then
    \[ \A\iid[v_s|Q] = \frac{\I[k > n-\lambda]}{n-\lambda+k},~~ \A\iid[v_d|Q] = \frac{\I[k \le n-\lambda]}{2h}. \]
    Similarly, we find the number of path traces $Q$ by finding the number of ways to put $k-1$ slots of black nodes into $h$ gaps, and $\lambda - 2h - k$ slots of white nodes into $h$ gaps. Hence,
    \[ \abs{\brs{Q \subset P}{v_s \in b_Q, v_d \in g_Q}} = \combus{h+k-2}{k-1} \combus{\lambda-h-k-1}{h-1}. \]
    
    \item If $v_s$ is gray and $v_d$ is black, by symmetry,
    \[ \A\iid[v_s|Q] = \frac{\I[k \le n-\lambda]}{2h},~~ \A\iid[v_d|Q] = \frac{\I[k > n-\lambda]}{n-\lambda+k}, \]
    \[ \abs{\brs{Q \subset P}{v_s \in g_Q, v_d \in b_Q}} = \combus{h+k-2}{k-1} \combus{\lambda-h-k-1}{h-1}. \]
    
    \item If both are gray, then 
    \[ \A\iid[v_s|Q] = \A\iid[v_d|Q] = \frac{\I[k \le n-\lambda]}{2h}, \]
    \[ \abs{\brs{Q \subset P}{v_s \in g_Q, v_d \in b_Q}} = \combus{h+k-2}{k} \combus{\lambda-h-k-1}{h-1}. \]
\end{enumerate}

The case $k < 2$ is not specified as $\combus{M}{N} = 0$ when $M < 0 \le N$. 

After the above discussion, we may conclude that for a particular $Q$, its term in the summation depends only on its $h$, $k$ and the color of $v_S$ and $v_d$. Therefore, the $2^\lambda$ terms in the original summation \eqref{eqn:prvpartp} are merged into $\mathrm{Poly}(n)$ terms as below.
\begin{align}
    & 1 - \prv(\D\iid) - \frac{2(1-\alpha)^L}{n} \notag \\
    =~& \sum_{h = 1}^{\floor{\lambda/2}} \sum_{k = 0}^{\lambda-2h} 
    \begin{aligned}[t] 
        & \alpha^{\lambda-h-k} (1-\alpha)^{h+k-1} \cdot \combus{\lambda-h-k-1}{h-1} \cdot 2 \cdot \\ 
        \bigg\{ & \combus{h+k-2}{k-2} \frac{\I[k > n-\lambda]}{n-\lambda+k}  \\
        + & \combus{h+k-2}{k-1} \brb{\frac{\I[k > n-\lambda]}{n-\lambda+k} + \frac{\I[k \le n-\lambda]}{2h}} \\
        + & \combus{h+k-2}{k} \frac{\I[k \le n-\lambda]}{2h}\bigg\}
    \end{aligned} \notag \\
    =~& 
    \begin{aligned}[t] 
        & \sum_{h = 1}^{\floor{\lambda/2}} \sum_{k = 0}^{\lambda-2h}  \alpha^{\lambda-h-k} (1-\alpha)^{h+k-1} \cdot \combus{\lambda-h-k-1}{h-1} \cdot 2 \cdot \\ 
        \bigg\{ & \combus{h+k-1}{k-1} \frac{\I[k > n-\lambda]}{n-\lambda+k} 
        + \combus{h+k-1}{k} \frac{\I[k \le n-\lambda]}{2h} \bigg\}
    \end{aligned} \notag \\
    =~& \sum_{h = 1}^{\floor{\lambda/2}} \sum_{k = 0}^{\lambda-2h} \alpha^{\lambda-h-k} (1-\alpha)^{h+k-1} \combus{\lambda-h-k-1}{h-1} \frac{\psi(k)}{h+k} \combus{h+k}{h} \notag \\
    =~& \sum_{t=1}^{\lambda-1} \frac{\alpha^{\lambda-t} (1-\alpha)^{t-1}}{t} \sum_{h=1}^{\min\brc{t, \lambda-t}} \combus{t}{t-h} \combus{\lambda-t-1}{h-1} \psi(t-h). \label{eqn:exactiid}
\end{align}

\eqref{eqn:exactiid} is the polynomial-time algorithm of computing $\prv(\D\iid)$. The tradeoff relationships of different settings of $(n, \lambda)$ is shown in Figure \ref{fig:perf}. Because of $\psi(t-h)$, the summation above is difficult to simplify. Thus, we apply $\psi(t-h) \le 2$ and provide a lower bound for privacy metric.
\begin{align*}
    & 1 - \prv(\D) - \frac{2(1-\alpha)^L}{n} \\
    \le~& 2\sum_{t=1}^{\lambda-1} \frac{\alpha^{\lambda-t} (1-\alpha)^{t-1}}{t} \sum_{h=1}^{\min\brc{t, \lambda-t}} \combus{t}{t-h} \combus{\lambda-t-1}{h-1} \\
    =~& 2\sum_{t=1}^{\lambda-1} \frac{\alpha^{\lambda-t} (1-\alpha)^{t-1}}{t} \combus{\lambda-1}{t-1} \\
    =~& \frac{2}{\lambda} \sum_{t=1}^{\lambda-1} \alpha^{\lambda-t} (1-\alpha)^{t-1} \combus{\lambda}{t} \\
    =~& \frac{2}{\lambda(1-\alpha)} \brb{1 - \alpha^\lambda - (1-\alpha)^\lambda}.
\end{align*}

The lower bound in \eqref{eqn:clq} is justified. \qedb

\subsection{Proof of Theorem \ref{thm:zerosr}}
\label{sec:pfzerosr}

Recall that the PCN is run as a Markov chain. In 3 steps, we prove the success rate tends to $0$ when the Markov chain becomes stable after infinitely many transactions.

\begin{enumerate}
    {\bf \item There exists an absorbing state. }
    
    For an edge $(u, v)$ with $0 \le b(u, v), \tilde b(v, u) < \ell$, it's impossible to route through this edge via any direction. So we call it {\bf locked}. 
    
    In $\gc$, we pick a set of edges $\ec' \subseteq \ec$ and set them locked, such that in the (undirected) graph $\gc' = \bra{\vc, \ec - \ec'}$, for any $(u, v)$ that is possible to be chosen as the pair of sender and recipient, $u, v$ belong to different connected components. Note that such $\ec'$ exists: $\ec'$ could be $\ec$, where all nodes are separated from each other.
    
    {\bf \item There's a non-zero probability that any transient state is transitioned to an absorbing state within finite steps.}
    
    In an transient state $\gc_0$ (excluding locked edges), there exists a pair of nodes $(u, v)$ which are still connected via some path $\pc$. 
    As assumed, there exists an edge $\epsilon \in \pc$ whose public balance is not updated with probability at least $\beta > 0$. 
    Assume it takes $k_1$ transactions of value $\ell$ to deplete $\epsilon$ on the direction to $v$. 
    After this, it takes another $k_2$ transactions of value $\ell$ to deplete $\epsilon$ on the opposite direction. 
    Assume probability of choosing $(u, v, \ell)$ and $(v, u, \ell)$ are $p_1$ and $p_2$, respectively. 
    Then the probability of locking $\epsilon$ after $k_1 + k_2$ transitions is at least $(\alpha p_1)^{k_1} (\beta p_2)^{k_2}$, which is positive. 
    
    Let $\gc_1$ denote the new state after locking $\epsilon$.
    Note that $\gc_1$ has one fewer edge than $\gc_0$. Since the number of edges is finite, there's a positive probability that $\gc_0$ is transitioned to a state where no pair of nodes is connected via a path of unlocked edges. 
    In this case, no more edge can be locked, and this is exactly an absorbing state discussed above. 
    
    {\bf \item Success rate reaches $0$ almost surely as $T\to\infty$.}
    
    Apparently, in any absorbing state, no transaction sampled from the same distribution can succeed. 
    In this Markov chain, at least one absorbing state is reachable from any transient state. 
    Thus, almost surely, the PCN will be at an absorbing state after finite number of transactions. 
    Because such PCNs will no longer be able to support any transaction sampled from the same distribution, the overall success rate is $0$. \qedb
    
    
    
    
    
    
\end{enumerate}

\section{Simulation Results}
\label{app:sims}

\subsection{Transaction Value}

\begin{figure}[ht]
    \centering
    \includegraphics[width=\mywidth]{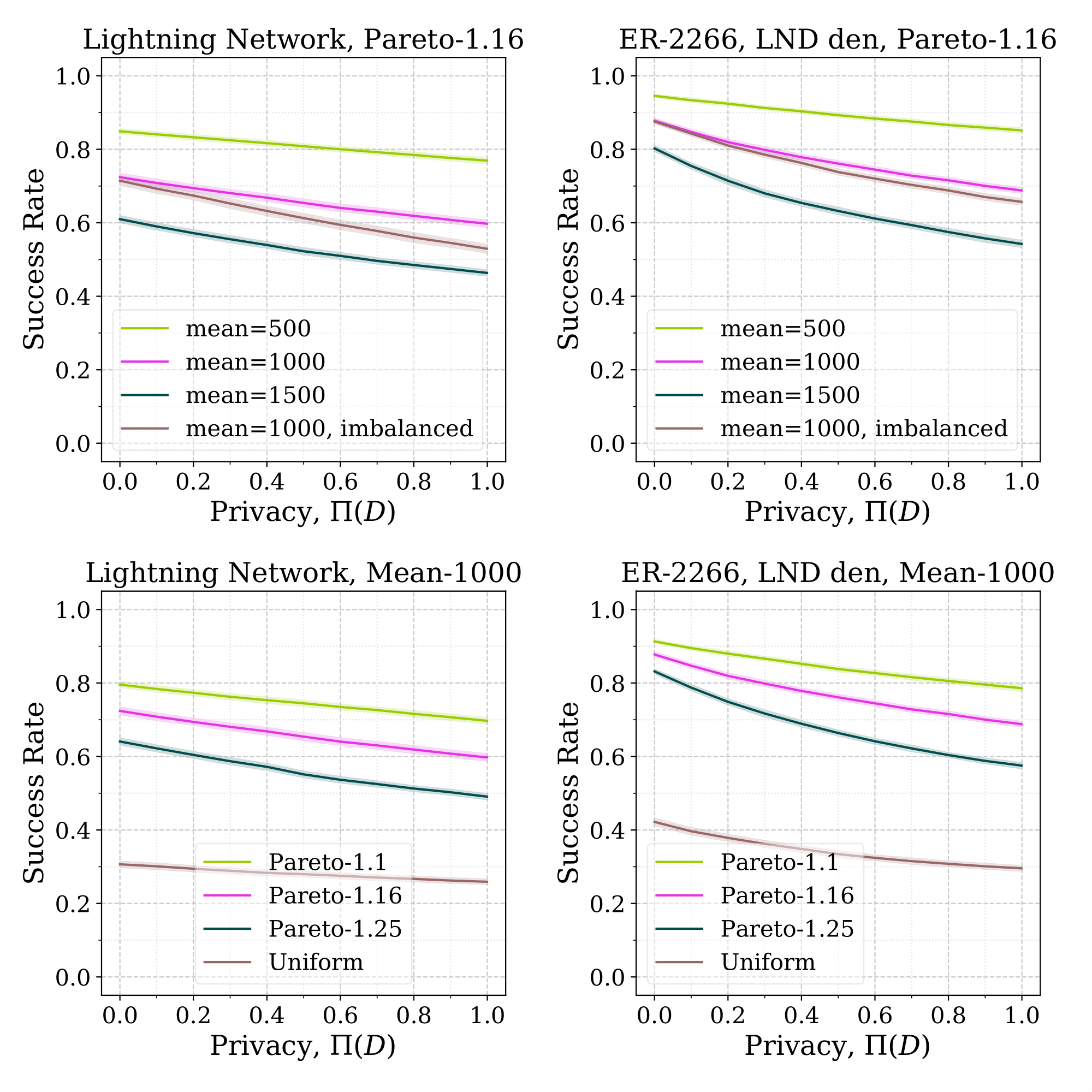}
    \caption{Success rate-privacy tradeoffs for different transaction value distributions. 
    The top row considers Pareto-distributed transactions of different means;  the bottom row considers different distributions with the same mean.}
    \label{fig:expval}
\end{figure}

In the experimental settings in Section \ref{sec:num}, the value of every transaction followed a Pareto distribution $\pareto(1.16, 1000)$. 
To study the curves for different transaction value distributions, we first selected two typical network structures: Lightning Network and \ert~ network with the same size and density. 
We also fixed the size of the workload to 3,000 transactions. 
We considered two properties of transaction values: distribution type and expectation. 
The distribution types include the uniform distribution from $0$ to twice the expected value, and 3 different Pareto distributions with $\beta = 1.1, 1.16, 1.25$. 
\apdx{Their PDFs are plotted in Figure \ref{fig:pareto} (Appendix \ref{app:sims}).}
Figure \ref{fig:expval} illustrates the privacy-success rate tradeoffs for each of these experimental settings. 
Here, the ``imbalanced'' curve in the top row refers to our approach for generating the endpoints of transactions; 
instead of choosing the source and destination uniformly, we assign users a weight of 1 with probability 0.8 and 16 with probability 0.2. 
When generating transactions, we sample endpoints with  probability proportional to their weights. 
On  average, 20\% of the users hold 80\% of the weight; this models the imbalance observed in real financial transactions. 
These plots show that when the mean transaction value increases, the success rate decreases. 
This is because we are keeping the capacity distribution fixed, so larger transactions are more likely to fail. 
Similarly, using increasingly heavy-tailed Pareto transactions (smaller $\beta$) causes the minimum transaction value to decrease, which increases the success rate.
Although these plots illustrate some trends related to transaction value distribution, the main takeaway is the fact that for a variety of transaction value distributions, the slope of the privacy-success rate curves is shallow. 
\wt{Figure \ref{fig:exptop} also shows the slopes of curves are also not decisively steep when the workloads are heavier.}
This implies that \textbf{one cannot trade small losses in privacy for large gains in success rate, or vice versa.}

\subsection{Workload size}
\label{sec:workload}
\begin{figure}[ht]
    \centering
    \includegraphics[width=\mywidth]{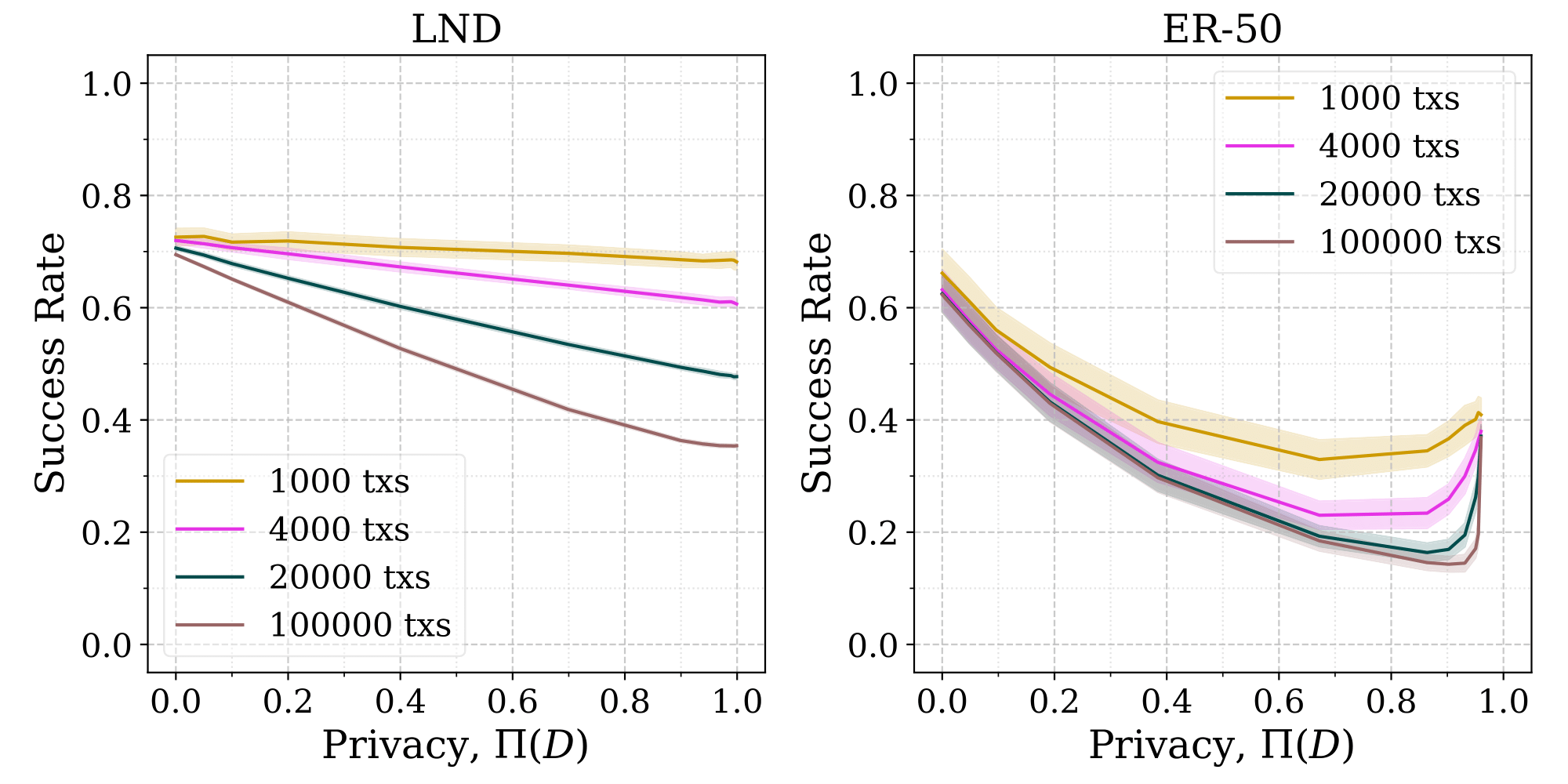}
    \caption{Transaction workload size vs success rate.}
    \label{fig:expwl}
\end{figure}

We next investigate the role of workload size by varying the number of transactions in our workload. 
Figure \ref{fig:expwl} shows the average success rate as a function of privacy as we process 
from 1,000 to 100,000 transactions. The ER graph has a density of $\log(50)/50$.
%
More precisely, this experiment revealed two unexpected phenomena: 
(1) for all privacy levels (including zero), success rate decreases as we process more transactions. 
It is not obvious \emph{a priori} why this should happen, as transactions are generated uniformly at random between nodes; 
on average, transaction flows should be balanced. 
(2) Success rate is not always monotonically decreasing in the privacy parameter.
\wt{This phenomenon is triggered by "deadlocks" introduced in Section \ref{sec:relation}. 
These deadlocks suggest that sacrificing a little privacy can actually \emph{hurt} average success rate in the long term. 
Hence system designers should be wary of deploying a noise mechanism that sacrifices only a little privacy; 
not only are the gains modest in theory, they can actually be negative in practice, unless system designers deal with deadlocks. }

\paragraph{Declining success rate over time}
The declining success rates are partially due to the previously-described deadlocks.
However, we observe the phenomenon even at zero privacy, when deadlocks cannot happen.
The reason may be related to the formation of \emph{routing bottlenecks}.
Suppose a node $v$ receives many transactions until all of $v$'s neighboring channels are imbalanced, with most of the balance on $v$'s side of the channel. 
In such a scenario, nobody can  route through $v$:
funds cannot reach $v$, so it cannot relay money.
Under randomized transaction workloads like ours, such bottlenecks are theoretically recurrent;
in practice, they can also occur if some nodes are popular money recipients (e.g., merchants).
The only way to break a bottleneck is for the bottlenecked node(s) to send their own transactions out of the isolated zone.
In a large network, the likelihood of this happening for a uniform workload is small. 
Hence, the system may be bottlenecked for a long time.

\subsection{Network Size}

To observe the effects of network size,
the source and destination  of each transaction are still drawn uniformly, and the values of transactions still follow Pareto distribution $\pareto(1.16, 1000)$. 
To generate smaller versions of the Lightning network, we snowball sample the full-sized graph until reaching the desired size. 
For each network size, we sample 100 networks, and the average density of both \ert~ and LND networks are matched.

\begin{figure}[H]
    \centering
    \includegraphics[width=\mywidth]{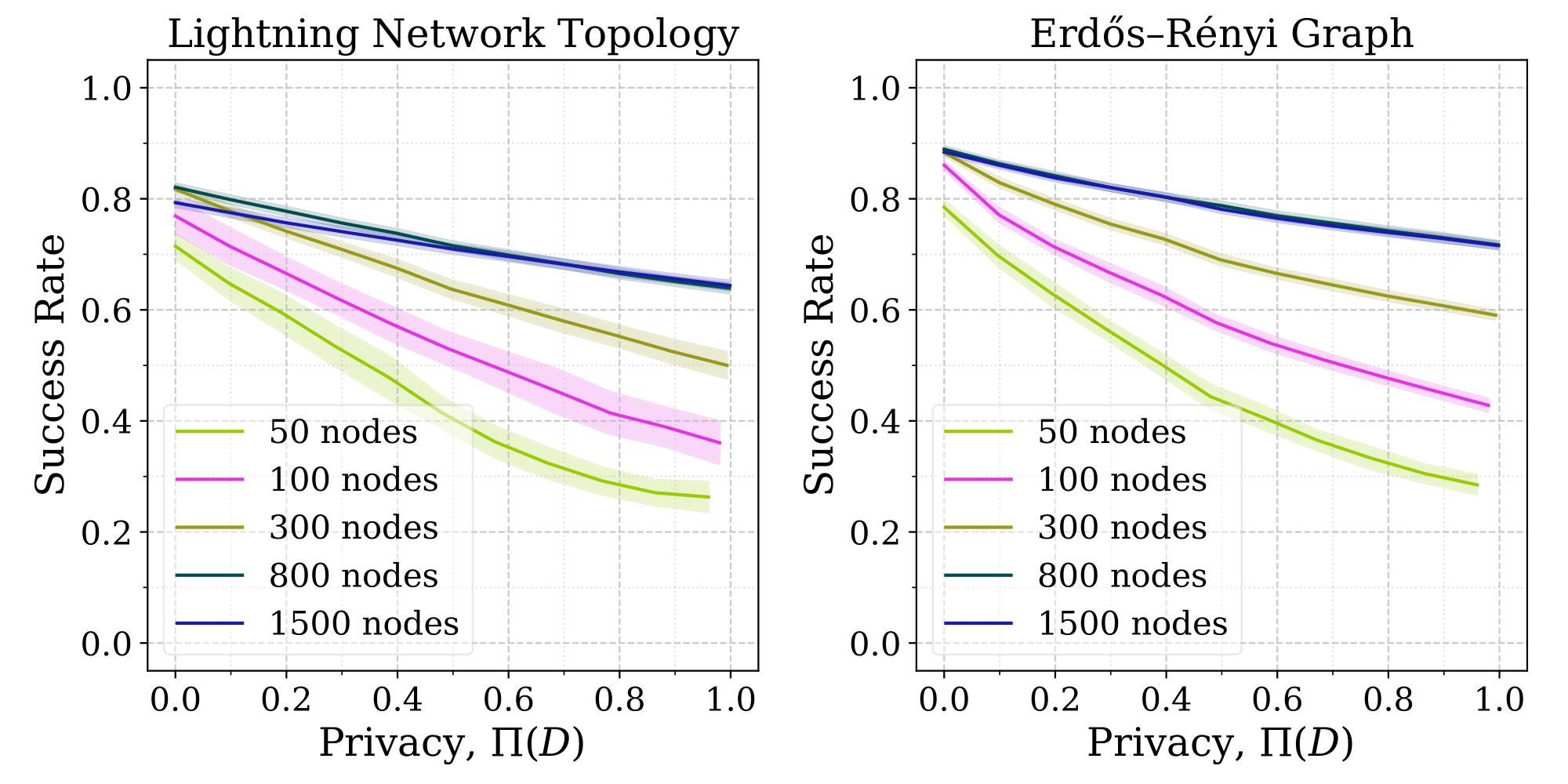}
    \caption{Privacy-Success rate tradeoff curves for different graph sizes.}
    \label{fig:expsize}
\end{figure}

The tradeoff curves are shown in Figure \ref{fig:expsize}. 
When the size is small, there is high variance in network topology of LND, which leads to high variance in success rates. 
As the network size increases, the privacy-success rate tradeoff becomes insensitive to these variations. 
Moreover, we observe that success rate increases. 
This may be because as more nodes are added, there are more paths for transactions to traverse, which gives additional robustness to link imbalance.


\subsection{Data distributions and workloads}
\begin{figure}[H]
    \centering
    \includegraphics[width=.9\mywidth]{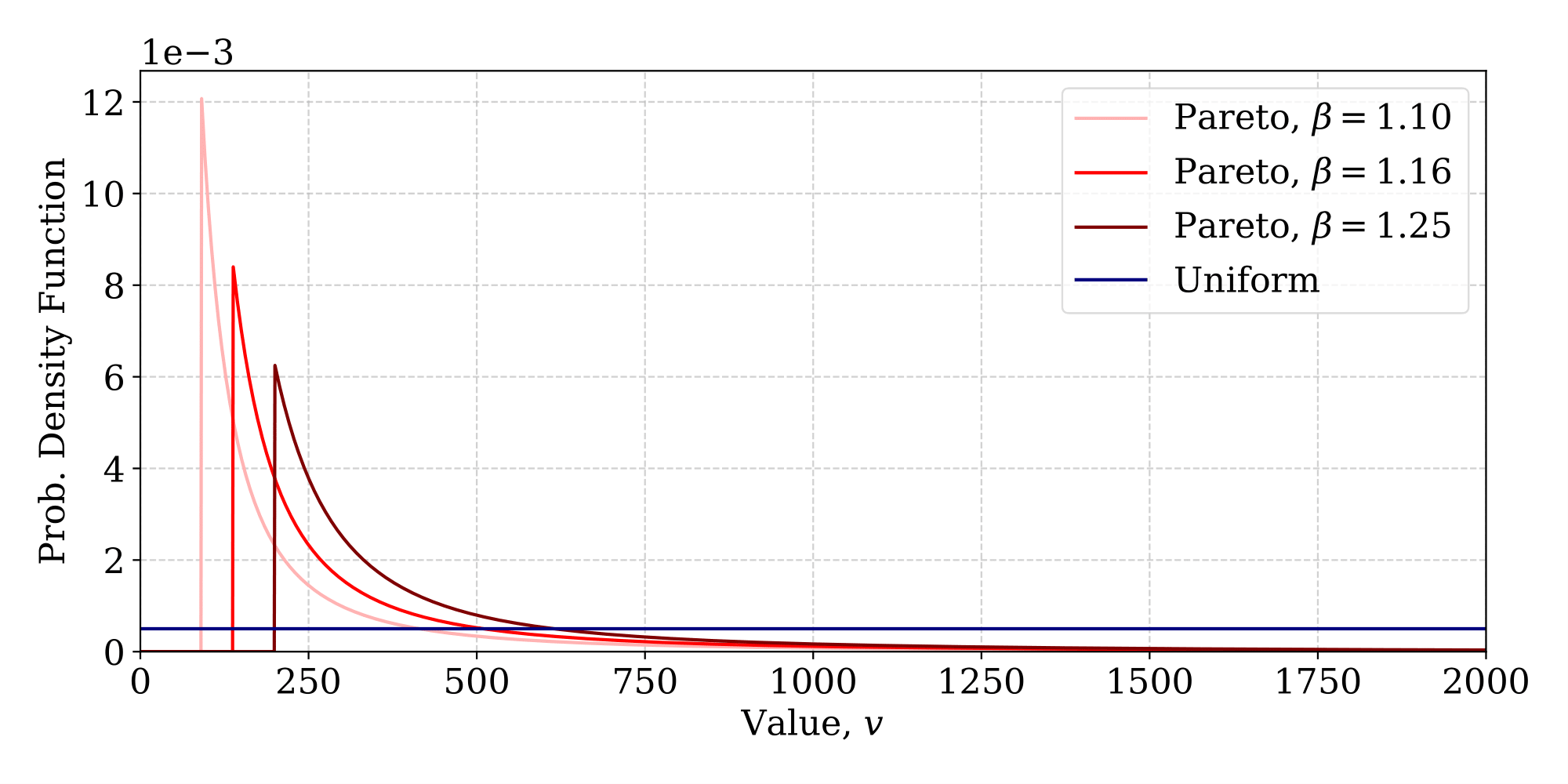}
    \caption{Pareto distribution pdfs for different parameter settings.}
    \label{fig:pareto}
\end{figure}

Figure \ref{fig:pareto} illustrates the pdfs of the Pareto distributions used in our experiments. 
Notice that Pareto distributions have a minimum transaction value, and are also heavy-tailed. 

\begin{figure}[H]
    \centering
    \includegraphics[width=3.2in]{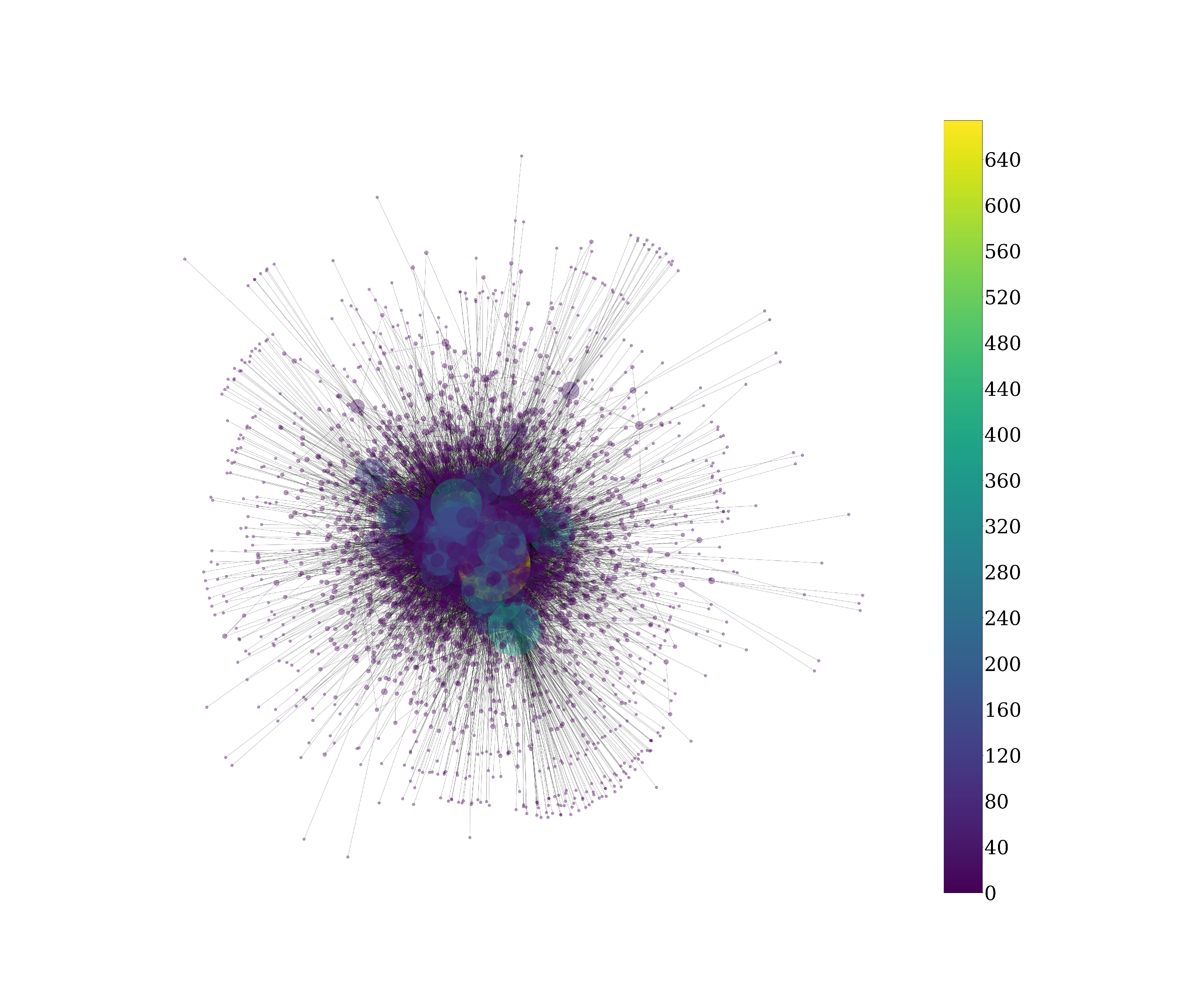}
    \caption{Snapshot of the Lightning network topology from December 28, 2018. 
    Node sizes are scaled proportionally to their degree in the graph.}
    \label{fig:lnd}
\end{figure}

Our experiments use a snapshot of the real Lightning network topology, illustrated in Figure \ref{fig:lnd}.
The graph has many nodes of degree 1 or 2, with several large hubs that are well-connected. 
This is qualitatively different from the synthetic ER graphs we also used in experiments. 
